\documentclass[preprint]{aastex62}
\usepackage[T1]{fontenc}
\usepackage[english]{babel}
\usepackage{epstopdf}
\usepackage{natbib}
\usepackage{ulem,xcolor}
\usepackage{amsmath, graphicx}
\usepackage{soul}
\def\rhessi{{\textit{RHESSI}}}
\def\goes{{\textit{GOES}}}

\def\kw{{Konus-\textit{Wind}}}
\def\sdo{{\textit{SDO}}}
\def\mw{{microwave}}
\def\Mw{{Microwave}}

\begin{document}

\title{Spatio-temporal energy partitioning in a non-thermally dominated two-loop solar flare}

\author[0000-0001-7856-084X]{Galina G. Motorina}
	\affil{Astronomical Institute of the Czech Academy of Sciences, 251 65 Ond\v{r}ejov, Czech Republic\\
Central  Astronomical Observatory at Pulkovo of Russian Academy of Sciences, St. Petersburg, 196140, Russia\\
Ioffe Institute, Polytekhnicheskaya, 26, St. Petersburg, 194021, Russia}

	\author[0000-0001-5557-2100]{Gregory D. Fleishman}
	\affil{Center For Solar-Terrestrial Research, New Jersey Institute of Technology, Newark, NJ 07102, USA}

	\author[0000-0002-8078-0902]{Eduard P. Kontar}
	\affil{School of Physics \& Astronomy, University of Glasgow, G12 8QQ,
Glasgow, United Kingdom}

\begin{abstract}
Solar flares show remarkable variety of the energy partitioning between thermal and nonthermal components. Those with a prominent nonthermal
component but only a modest thermal one are particularly well suited to study the direct effect of the nonthermal electrons on plasma heating.
Here, we analyze such a well observed, impulsive single-spike nonthermal event,
a SOL2013-11-05T035054 solar flare,
where the plasma heating can be entirely attributed to the energy losses of these impulsively accelerated electrons. Evolution of the energy budget of thermal and nonthermal components during the flare is analysed using X-ray, microwave, and EUV observations and three-dimensional modeling.
The results suggest that (i) the flare geometry is consistent
with a two-loop morphology and the magnetic energy is likely released due to interaction between these two loops;
(ii) the {released} magnetic energy converted to the nonthermal energy of accelerated electrons only,
which is subsequently converted to the thermal energy of the plasma; (iii) the energy is partitioned in these two flaring loops in comparable amounts;
(iv) one of these flaring loops remained relatively tenuous but rather hot,
while the other remained relatively cool but denser than the first one.
Therefore, this solar flare demonstrates an extreme efficiency of conversion of the free magnetic energy to the nonthermal energy of particle acceleration and the energy flow into two loops from the non-thermal to thermal one with a negligible direct heating.
\end{abstract}

\keywords{Sun: Flares - Sun: X-rays, EUV, Radio emission}

\section{Introduction}
\label{S_Intro}

Magnetic energy accumulated in the solar atmosphere over days can be sporadically promptly released on the time scale of minutes to produce a solar flare---a transient brightening observed virtually across the whole electromagnetic radiation spectrum from radio to gamma rays.
The current understanding of solar flares \citep[e.g. ][as a review]{2011SSRv..159..107H} suggests that the primary energy release due to magnetic reconnection produces a downward beam of accelerated electrons.
These energetic electrons generate nonthermal microwave gyrosynchrotron (GS) radiation in the ambient magnetic field and {hard X-ray} (HXR) emission
in interactions with ambient particles.
These collisions with plasma result in heating of the coronal plasma
and dense chromospheric layers stimulating evaporation of the heated chromospheric plasma into the coronal portion of the flaring loop(s). Finally,
the hot plasma cools down due to conductive and radiative losses,
whose signatures are routinely observed at {soft X-ray} (SXR)
and ultraviolet (UV) wavelengths.

The close causal relationship between nonthermal emissions,
produced by accelerated particles, and thermal emissions,
produced by the plasma that is heated during the flare
and cools down after the energy release,
is one of the major ingredient for the standard model {of solar flare \citep[CSHKP:][]{1964NASSP..50..451C, 1966Natur.211..695S,  1974SoPh...34..323H, 1976SoPh...50...85K, 1995ApJ...451L..83S, 1997ApJ...483..507T, 1999Ap&SS.264..129S}}.
Such a casual relationship, known as the Neupert effect \citep{1968ApJ...153L..59N},
is often observed in flares \citep{1975IAUS...68..191D,1993SoPh..146..177D,1993A&A...267..586S,1999ApJ...514..472M}.
In many cases, however, the Neupert effect is not observed \citep{1993SoPh..146..177D,2005ApJ...621..482V, 2008AdSpR..41..988S, 2009A&A...498..891B},
which indicates richer physics of plasma heating than the simple physics of the Coulomb
collisional loss by nonthermal particles.

Accordingly, a number of physical mechanisms for a `direct' heating
of coronal plasma have been proposed such as adiabatic heating due to Fermi
\citep{1990ApJ...361..701M,1996AIPC..374....3C,2017PhPl...24i2110D}
and betatron mechanisms \citep{1965PASJ...17..403S,1975ApJ...200..734B,1979SvAL....5...28S,2005AstL...31..537B, 2012A&A...546A..85G},
due to reconnecting current sheet located at the apex of the magnetic cusp,
due to shock waves associated with the outflow of plasma from the current sheet  \citep[see, e.g.,][]{2002A&ARv..10..313P},
or due to strong electric fields that might be produced by the Rayleigh-Taylor instability \citep{2016Ge&Ae..56..952S, 2017SoPh..292..141Z}.
In addition, the heat-transport is likely to be complicated by non-local
effects due to electrostatic, magnetostatic and/or electromagnetic
turbulence \citep{1979SvAL....5...28S,1991AdSpR..11...99J,1998A&A...334.1112J,2016ApJ...824...78B}

As a result, the partitions between the thermal and nonthermal energies vary
in extremely broad range in solar flares---from entirely thermal \citep{1985SoPh..100..465D,Gary_Hurford_1989, Altyntsev_etal_2012,Fl_etal_2015} to essentially nonthermal \citep{1985SoPh..100..465D,2007ApJ...666.1256B, 2010ApJ...714.1108K, 2011ApJ...731L..19F}.
In spite of this variety of the thermal-to-nonthermal energy ratios,
the flare magnitude is commonly characterized by
SXR class based on the flux
of the thermal SXR emission. Should the thermal SXR flux be somewhat low,
the flare is deemed weak even though it may contain a strong nonthermal component, which, in some cases, can be dominant
in the flare energy budget.

Intuitively, such nonthermal-dominated events should demonstrate the Neupert effect; however, the opposite is not necessarily true: the presence of the Neupert effect does not guarantee the dominance of the nonthermal energy in the flare.
For example, \citet{2005ApJ...621..482V} found that the correlations between \textit{physical values} of the thermal and nonthermal \textit{energies}
were lower than correlations between \textit{phenomenologically} defined thermal and nonthermal \textit{emissions}, which was interpreted as evidence of additional `direct' plasma heating. Thus, the nonthermal-dominated flares represent, likely, only a subset of impulsive flares showing the Neupert effect. The criticle obstacle
in determining non-thermal energy and hence the flare power is related to poorly known low-energy cut-off in the energy spectrum of accelerated particles. Low energy cut-off problem arises because of the low-energy part of electron spectrum is often masked by bright thermal emission. Thus, this low-energy part of the nonthermal electron spectrum could be much better studied in flares with anomalously weak thermal response compared with a ``normal'' flare. In addition,
recent development of the warm-target model allowed  \cite{2019ApJ...871..225K} to obtain the low-energy electron cutoff with $\sim7\%$ statistical uncertainty at the $3\sigma$ level for a selected flare, hence,
allowing quantitative studies of the non-thermal energetics in solar flares.

For non-thermal electron analysis, it is advantageous to use the nonthermal-dominated events within those that does not show any appreciable preflare heating but demonstrate a prominent impulsive nonthermal phase,
which is followed by thermal emission, called the `early impulsive flares'  \citep{1994ApJ...421..843F,2007ApJ...670..862S}.
It has recently been recognized that some early impulsive flares \citep{1992ApJ...384..656W, 2007ApJ...666.1256B, 2011ApJ...731L..19F, 2013PASJ...65S...1M, 2016ApJ...822...71F} have only a very little thermal plasma response so that these events are often even not listed as \goes\ flares. For that reason, they were classified as `cold' flares \citep{2018ApJ...856..111L}. However, the 2002-Jul-30 `cold' flare reported by \cite{2011ApJ...731L..19F}
is a tenuous flare with the thermal number density not exceeding $2\times10^9$ cm$^{-3}$ at the coronal part of the flaring loop.
Therefore, the thermal SXR emission is low because the emission measure is low, although the temperature can in fact be rather large. On the contrary,
two other cold flares reported by \cite{2007ApJ...666.1256B} and \cite{2013PASJ...65S...1M} were dense, with a thermal number density in excess of $10^{11}$ cm$^{-3}$. In such cases, the fast particle losses in the coronal part of the loop are large and the thermal energy increase is relatively strong;
but, because of the high density, the net temperature increase,
and hence the emission of SXR, is rather modest.
However, previous studies of such events do not provide a comprehensive picture of energy assessment due to weak thermal component being poorly constrained from \goes\ observations only \citep{2007ApJ...666.1256B, 2011ApJ...731L..19F, 2016ApJ...822...71F}.

Recently, a statistical study of the `cold' flares has been performed by \cite{2018ApJ...856..111L} for the period from 1994 to 2017 with 27 selected events, where the authors classified such events as short and hard flares,
which are typically produced in compact structures (likely, short flaring loops)
with strong magnetic field, stronger compared with the `normal' flares.
\citet{2018ApJ...856..111L} proposed that these `cold' flares may offer a rather clean case to study the thermal plasma heating in response to the nonthermal electron acceleration in flares.
Indeed, for such events the spatially-resolved thermal energy evolution
can be quantified in detail using EUV data, since in the case of relatively modest plasma heating, the EUV images are not affected by saturation
and the flare plasma temperature is expected to be in the range
to which \sdo/AIA is most sensitive. Thus, \sdo/AIA observations are much better suited to quantify the thermal component than spatially-unresolved \goes\ data employed so far \citep{2007ApJ...666.1256B, 2015ApJ...815...73B, 2016ApJ...822...71F, 2018ApJ...856..111L, 2019ApJ...872..204B}.

The present paper is focused on one of the events {(SOL2013-11-05T035054)} from the \cite{2018ApJ...856..111L} list, which was well-observed with a right set of imaging and spectroscopic instruments. Here, for the first time, we employ a detailed analysis of EUV data from \sdo/AIA to quantify the moderately heated component of the flare plasma, while use the \rhessi\ data to quantify the nonthermal component along with the hot component of the flaring plasma.
Combining these data with three-dimensional (3D) modelling based on \sdo/HMI vector magnetogram and nonlinear force-free field (NLFFF) reconstruction, we devise a 3D model of this flare and validate it by comparison with X-ray and \mw\ data.
This analysis reveals two distinct flaring loops,
one of which being indeed relatively cold, $\lesssim$10~MK. The other one, however, is rather hot, with a temperature up to $\sim$30~MK during the impulsive phase. Furthermore, the evolution of the spatially-resolved thermal energy is governed by the nonthermal component of the flare.
We found that the thermal energy evolution is quantitatively consistent with the rate of the nonthermal energy deposition; thus, here we do have the case, where the flare energy goes primarily to the particle acceleration, while the observed plasma heating is entirely accounted by dissipation of this nonthermal energy; there is no room for any statistically significant additional `direct' plasma heating in this event.

\section{Observations} \label{S_Observations}
Solar flare SOL2013-11-05T035054 occurred at AR 11890 at 03:48:40~UT, had the position (-771", -250") in heliocentric coordinates, localized by NoRH \citep[Table 1, ][]{2018ApJ...856..111L}. It demonstrated an impulsive HXR peak at energies above 12~keV that reached a maximum at $\sim$03:50:25~UT, while a smoother peak at lower energies that reached a maximum at 03:50:54~UT and ended at 03:56:40~UT based on the \rhessi\ observations.
The flare was included at the list of early impulsive `cold' flares studied by \cite{2018ApJ...856..111L}, who analysed soft and hard X-ray and microwave emission, observed with Geostationary Operational Environmental Satellite \citep[\goes ,][]{2005SoPh..227..231W}, \kw\ \citep{1995SSRv...71..265A, 2014Ge&Ae..54..943P}, Nobeyama Radio Polarimeters \citep[NoRP,][]{1979PRIAN..26..129T}, Radio Solar Telescope Network  \citep[RSTN,][]{1981BAAS...13Q.553G}, and Solar Radio Spectropolarimeters \citep[SRS,][]{Muratov2011}.
Here, for a detailed case study of the SOL2013-11-05T035054 flare we additionally employ data from Reuven Ramaty High Energy Solar Spectroscopic Imager \citep[\rhessi ,][]{2002SoPh..210....3L}, which provides a  few  arcsecond  angular resolution images and photon spectra at energies from $\sim$3 keV to 17 MeV with modest temporal resolution (4 s). We also use data obtained with Atmospheric Imaging Assembly on board Solar Dynamics Observatory  \citep[\sdo/AIA,][]{2012SoPh..275...17L} to more accurately quantify thermal and nonthermal energies of the solar flare.

\begin{figure}\centering
\includegraphics[width=12cm]{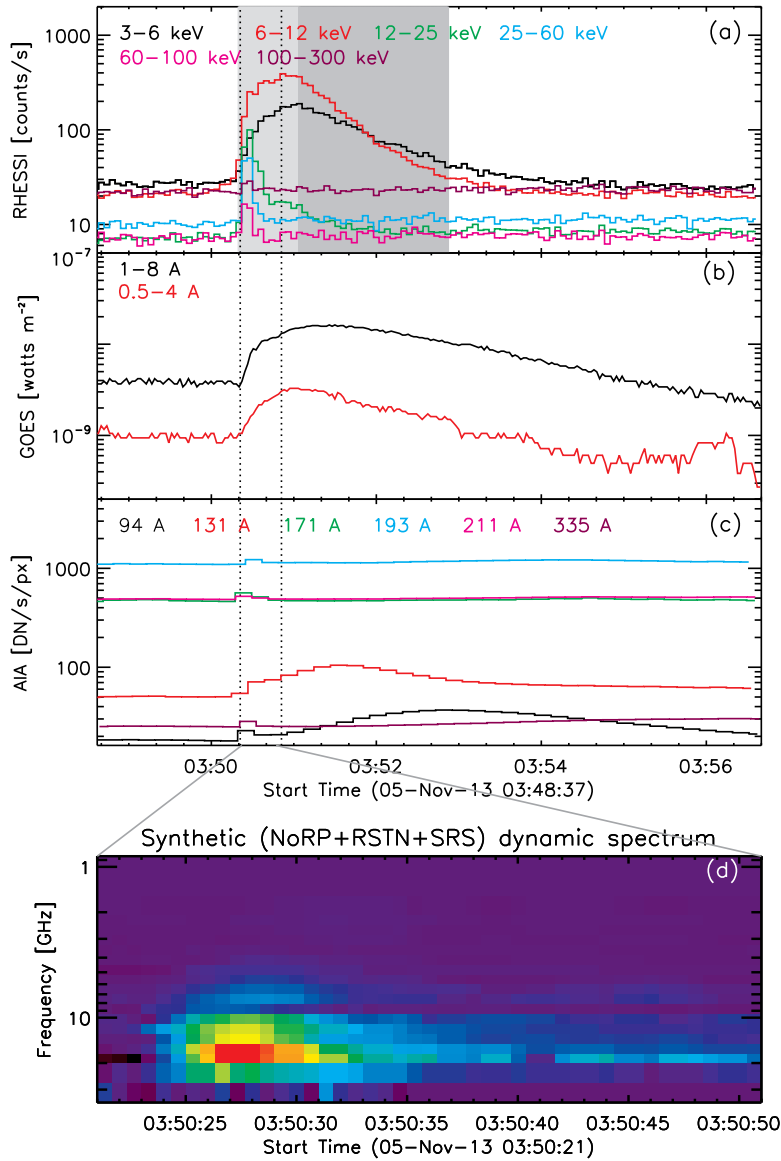}
\caption{Overview of the November 5, 2013 flare. From upper to bottom panel: {\rhessi\ light curves (a); \goes\ light curves (b): to highlight the flare-associated flux at the
1-8 \AA\ light curve, we subtracted a substantial fraction  $8\times 10^{-7}$~[watts m$^{-2}$] of the pre-flare background; the AIA light curves obtained from the selected FOV (see Fig.~\ref{Fig_AIA_map}) (c); NoRP$+$RSTN$+$SRS dynamic spectrum (the absolute peak of the radio flux density is 318\,sfu) of the impulsive flare phase (d).} Vertical dotted lines indicate the impulsive phase shown in the bottom panel. The light gray area shows the 4s fitted time intervals, the dark gray---the 12s fitted time intervals of the \rhessi\ observations.
\label{Fig_lightcurves}
}
\end{figure}

The flare has a short impulsive phase, $\sim$15 seconds, prominently seen in the nonthermal microwave and HXR emissions, and followed by $\sim$3 minute gradual phase, thus clearly showing the Neupert effect \citep{1968ApJ...153L..59N}.
The flare produced strong impulsive emission in HXR (up to 300 keV), microwave, and EUV wavelength domains seen,  respectively, by \rhessi\ and \kw, NoRP, RSTN, SRS, and \sdo/AIA at $\sim$03:50:20~UT, while a very modest SXR enhancement.
Figure \ref{Fig_lightcurves} {shows \rhessi\ (panel (a)) and \goes\ (panel (b)) light curves, EUV light curves (panel (c)) integrated over the field of view (FOV) shown below in Figure \ref{Fig_AIA_map} obtained with \sdo/AIA, and a synthetic microwave spectrum combining the data of the impulsive flare phase obtained with NoRP, RSTN, and SRS (panel (d)).} The radio spectral peak frequency was unusually high ($\sim$20 GHz) and the microwave burst was very short $\sim$15 s, which indicates  a reasonable compactness of the flare source.
The \sdo/AIA images were not saturated during the entire flare despite the presence of nonthermal emission up to 300~keV. The maximum of X-ray emission for $E>12$ keV and the microwave emission occurs simultaneously with bumps at  EUV light curves at $\sim$03:50:25~UT, while EUV peaks at 94, 171, 193, 211, 335 \AA\ point out an impulsive heating process for low temperatures ($\log_{10}T\sim$5.5-6.8) and the following cooling.
Summarizing, the described flare had a single impulsive peak followed by a reasonably simple, modest thermal response and it can be considered as an 'elementary' energy release/acceleration episode.

\subsection{EUV and X-ray imaging  with AIA and \rhessi\ }\label{S_imaging}

\begin{figure}\centering
\includegraphics[width=0.9\columnwidth]{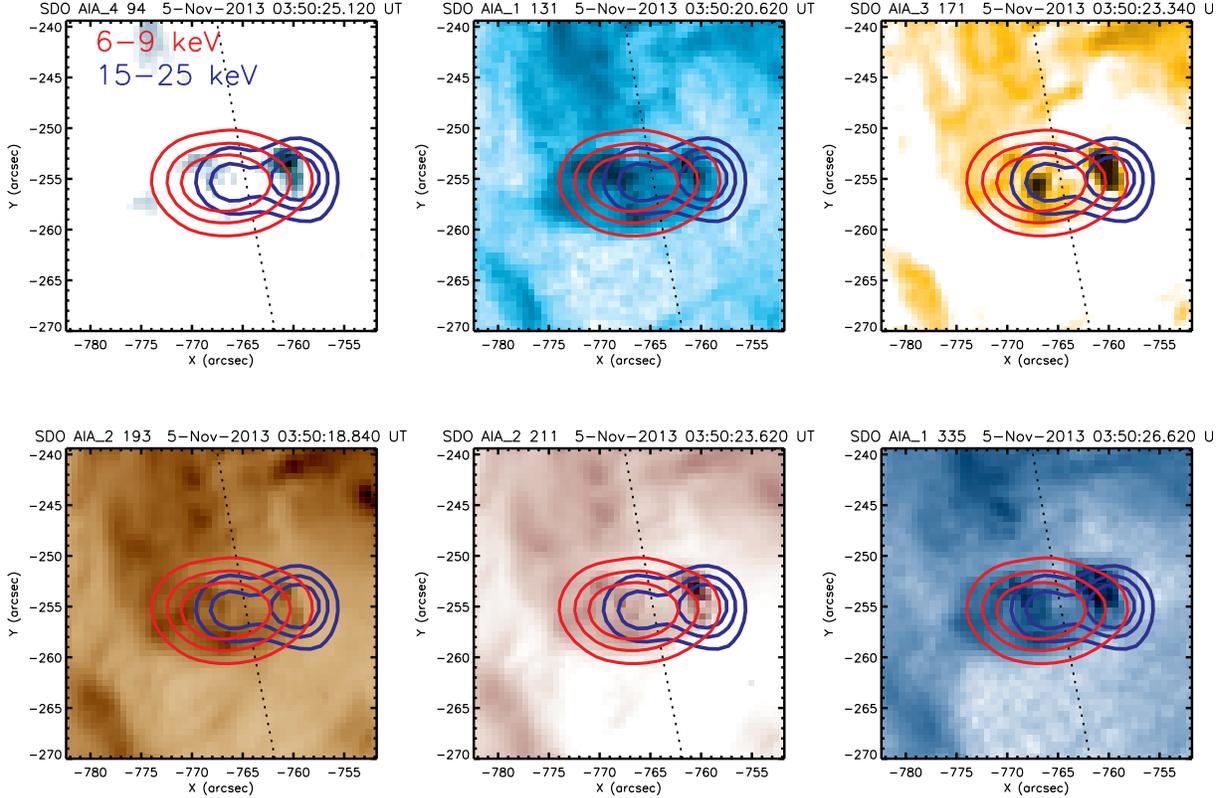}
\caption{AIA maps of the November 5, 2013 flare
with overlaid \rhessi\ CLEAN image
({Clean Beam Width Factor is} CBWF$=1.8$) 30, 50, and 70\% contours for 6-9 keV (red lines) and 15-25 keV (blue lines) for time interval 03:50:20-03:50:40~UT. {For co-alignment of the \rhessi\ and AIA maps, the AIA roll angle has been selected at AIA\_roll\_angle$=0.11$.}
\label{Fig_AIA_map}
}
\end{figure}

\sdo/AIA {observations provide the EUV full solar disk images} with a $\sim1.5\arcsec$ spatial  and 12\,s temporal  resolution  in several wavebands. In this study only six EUV wavebands (94, 131, 171, 193, 211, 335 \AA) are used, which are sensitive to emission of a coronal flare plasma. The AIA images additionally calibrated with aia\_prep.pro and normalized by the exposure time in 6 EUV wavebands were taken each 12\,s at the time interval between 03:48:37 and 03:56:40~UT. There was no saturation for the AIA data during the flare.
To analyse the EUV data we choose a square region that fully inscribes the flare emission at the entire course of the flare (see the FOV shown in Figure \ref{Fig_AIA_map}). The AIA Data Numbers (DN) were inferred from that FOV, which will be further investigated in Section\,\ref{S_aia_diagn}.

Figure\,\ref{Fig_AIA_map} also shows \rhessi\ CLEAN image as 30, 50, and 70\% contours for 6-9 keV and 15-25 keV for 03:50:20-03:50:40~UT time interval.
{To co-align the AIA (spatial resolution $\sim1.5\arcsec$ )  and \rhessi\ ($\sim2.5\arcsec$) observations the AIA maps were rotated (AIA\_roll\_angle$=0.11$) due to probable roll-angle calibration error in one of the instruments as earlier shown by \cite{2011A&A...533L...2B} and further discussed by \cite{2016ApJ...816....6K}.}
The AIA images at 94 and 171 \AA\ display double footpoints that lie inside the HXR contours. This likely points at the existence of two connecting loops in the horizontal direction. Loop-loop interactions have been proposed by \citet{1987ApJ...321.1031T,1999PASJ...51..483H} to drive energy release in solar flares with multi-loop morphology; this scenario is further supported by simulations of loop-loop interaction \citep{2005SoPh..232...63K} and numerous observations of two-loop flare configurations \citep[e.g.,][]{2016ApJ...822...71F,2019ApJ...883...38A}.

To determine the thermal and nonthermal X-ray sources the \rhessi\ CLEAN  \citep{2002SoPh..210...61H} images for 6-9 keV and 15-25 keV {with Clean Width Beam Factor CBWF=1.8} \citep[cf.][]{Kontar_2010}, {using detectors 3-8 with $\sim 6.79\arcsec$ nominal spatial resolution} were made for 03:50:20-03:50:40~UT time interval, which corresponds to the HXR impulsive peak (Figure \ref{Fig_AIA_map}, see Section\,\ref{S_aia_diagn_integral}).

\subsection{X-ray diagnostics with \rhessi\ and \kw}\label{S_rhessi_diagn}
To analyze the nonthermal properties of the flare, we employ spectral analysis in the X-ray domain.
The \rhessi\ data with appropriately subtracted background were fitted with two models: (i) isothermal plus collisional thick-target model and (ii) isothermal plus warm-target model {using OSPEX\footnote{For OSPEX documentation see \url{https://hesperia.gsfc.nasa.gov/rhessi3/software/spectroscopy/spectral-analysis-software/index.html}}}.
These fits were applied each 4s from 03:50:20~UT to 03:51:04~UT as this interval is of the greatest interest and each 12s for 03:51:04-03:52:52~UT time interval (the intervals are respectively highlighted by the light and dark gray areas in Fig.\ref{Fig_lightcurves}a), when the signal in the low-energy (thermal) channels was much higher than the background. Two examples of the \rhessi\ fits are shown in Fig.\ref{Fig_RHESSI_fits}, which illustrates a transition from nonthermal- (left panel) to thermal-dominated (right panel)  phases a few seconds after the HXR peak at $\sim$03:50:20~UT.
Starting $\sim$03:50:36~UT,  the nonthermal component drops below the background and the thermal component starts to dominate. The time evolution of the fit parameters is shown in Figure \ref{Fig_EM_T}: the temperature $T_{\rm{RHESSI}}$ {(a)} and the emission measure $EM_{\rm{RHESSI}}$ {(b)}, the total integrated electron flux $F_0$ [$10^{35}$ electrons/s] {(c)}, the low-energy cut-off $E_{c}$ [keV] {(d)}, and the spectral index  $\delta$ of the electron distribution function above $E_c$ {(e)}.

{It is important to emphasize here, that the physical meaning of the low-energy cut-off fit parameter $E_{c}$ is different in the cold-target and warm-target models: it gives only an upper limit of the ``true'' low-energy cut-off in the actual value in the spectrum of the nonthermal electrons in the case of cold target, while yields the most likely (mean) value in case of the warm-target model  \citep{2019ApJ...871..225K} given the adopted parameters of the coronal flaring loop. It is remarkable, however, that for this flare the upper limits (black symbols in panel d) and the mean values (red symbols in panel d) perfectly match each other within the statistical uncertainties; thus, the upper limits can safely be used as a proxy for $E_c$. This is important because other warm-target fit parameters (not shown in the figure) come with larger uncertainties than the thick-target parameters. Thus, in what follows we employ the better constrained parameters of the cold thick-target fit.
}

In addition, to probe if any fast, sub-second dynamics of the  nonthermal component is present, we employed the X-ray data obtained with the \kw\
\citep{1995SSRv...71..265A, 2014Ge&Ae..54..943P}. Given that the \kw\ uses triggering and adaptive spectrum accumulation scheme,
only four spectra with 64~ms temporal resolution were available for the 5 November 2013 flare for the time interval 03:50:24.587-03:50:24.843~UT during the rise of the impulsive phase.
We fitted the \kw\ data with isothermal plus thick target model, where thermal parameters were taken from the \rhessi\ fit and fixed as \kw\ registers photons in energy channels above $\sim$20 keV and thus is insensitive to thermal flaring plasma. The low energy cut-off $E_c=10.26$ keV was taken from values obtained from the \rhessi\ thick-target fit for the same time interval 03:50:24-03:50:28 UT. This fit does not reveal any significant time variability at the scale of $\sim$100~ms. Therefore, we added up all four spectra and fitted this cumulative \kw\ spectrum to cross-check the results of the \rhessi\ fit.
The total integrated electron flux and spectral index from the \kw\ fit are shown in Figure \ref{Fig_EM_T} (blue circles, {panels (c, e)}) and are consistent with the  fit parameters obtained from the \rhessi\ data.

\subsection{Thermal plasma diagnostics with \sdo/AIA}\label{S_aia_diagn}
\subsubsection{Diagnostics using the regularized DEMs}
\label{S_aia_diagn_integral}
To better quantify the thermal properties of the SOL2013-11-05T035054 flare, the observations of the \sdo/AIA were employed.
We note that \sdo/AIA is only weakly sensitive to the plasma much hotter than 10~MK \citep{2015Ge&Ae..55..995M, 2019ApJ...872..204B} and also have a gap in the sensitivity around 5\,MK; thus, we are only going to employ the AIA data to quantify the relatively cool component ($\sim$ 10~MK) of the `cold' flare plasma in addition to the much hotter component ($\sim$ 20--30~MK) already quantified with the \rhessi\ data. We also note that various methods of DEM reconstruction from the AIA data, or even the same method with different settings, may yield dissimilar DEM distributions; however, the moments computed from these distributions (i.e., emission measure and temperature) are more robust than the DEMs themselves. Therefore, we will focus on these integral parameters---emission measure $EM_{\rm{AIA}}$ [cm$^{-3}$] and mean temperature $\langle T_{\rm{AIA}} \rangle$ [K].

To estimate these integral plasma parameters from the AIA FOV with the area $A$ [cm$^{2}$] (see Fig.\ref{Fig_AIA_map}), we determined the differential emission measure (DEM) $\xi(T)$ [cm$^{-5}$K$^{-1}$] in the temperature range 0.5--25\,MK using a regularization technique applied to the entire FOV for the AIA data \citep{2012A&A...539A.146H}. The total emission measure from the AIA FOV can be computed as:
\begin{equation}\label{eq1}
EM_{\rm{AIA}}=A\int\limits_{T_{\min}}^{T_{\max}} \xi(T) dT \approx A\sum_i  \xi_i (T_i) \Delta T_i 
\end{equation}
and then the mean temperature as:
\begin{equation}\label{eq2}
\langle T_{\rm{AIA}} \rangle  =\frac{\int\limits_{T_{\min}}^{T_{\max}} T \xi(T) dT}{\int\limits_{T_{\min}}^{T_{\max}} \xi(T) dT} \approx \frac{\sum_i T \xi_i (T_i) \Delta T_i}{\sum_i  \xi_i (T_i) \Delta T_i}.
\end{equation}
{Figure \ref{Fig_EM_T}a,b shows evolution of $\langle T_{\rm{AIA}} \rangle$ and $EM_{\rm{AIA}}$ with the minimum preflare emission measure subtracted.
After the HXR impulsive peak, the emission measure and temperature both increase; however, the values obtained, respectively, from \rhessi\ and \sdo/AIA, are different from each other.
Due to sensitivity to different temperature ranges, the two instruments see different sources within the flaring region until $\sim$03:51:46 UT. After that, both $T$ and $EM$ obtained from \rhessi\ and \sdo/AIA converge, which likely indicates that the originally hotter plasma component, initially observed with \rhessi, has cooled down with time and became undetectable.
Finally, at the late decay phase, \rhessi\ becomes insensitive to the cooling plasma and only AIA could fully observe the decay phase of the flare.

\begin{figure}\centering
\includegraphics[width=8cm]{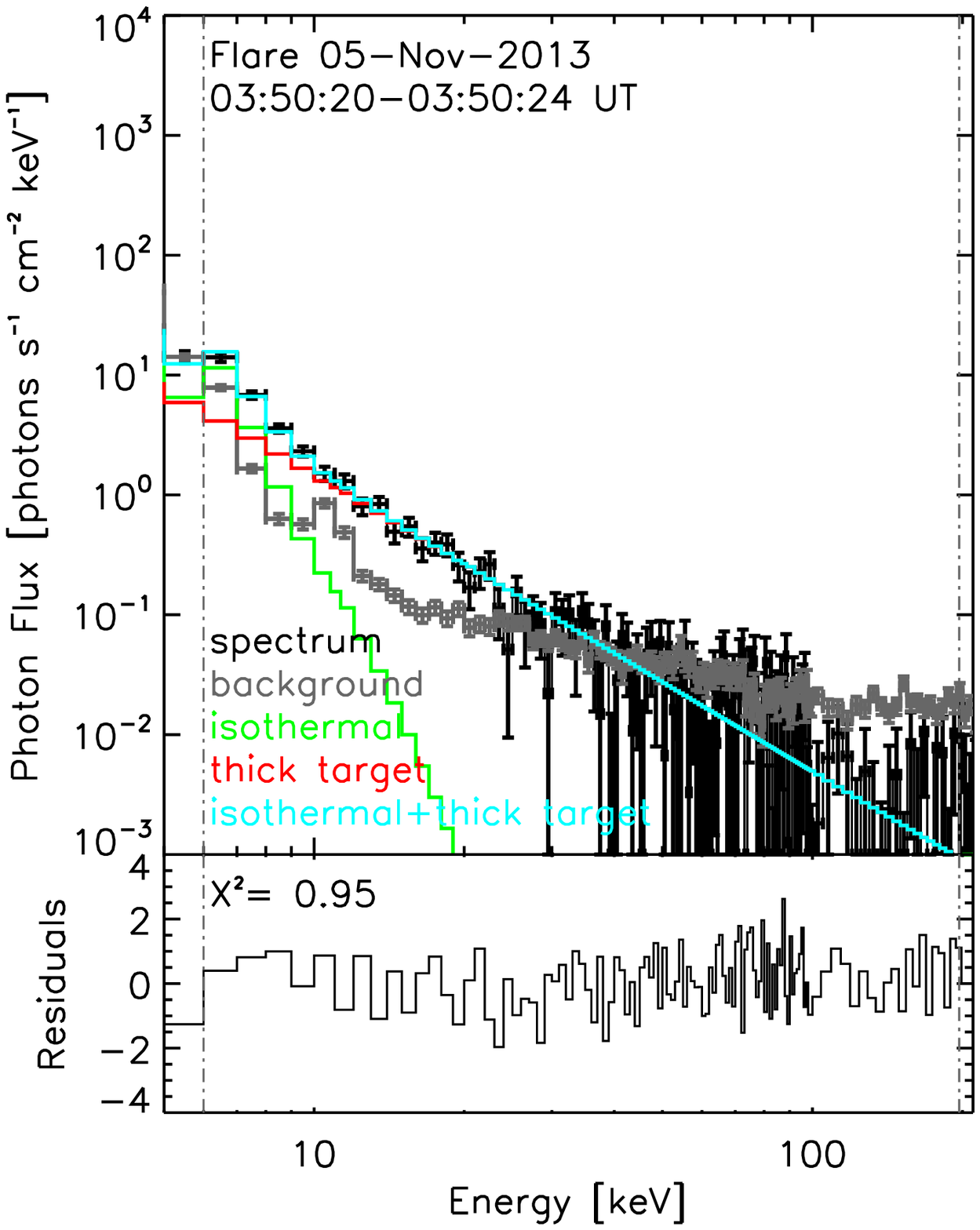}
\includegraphics[width=8cm]{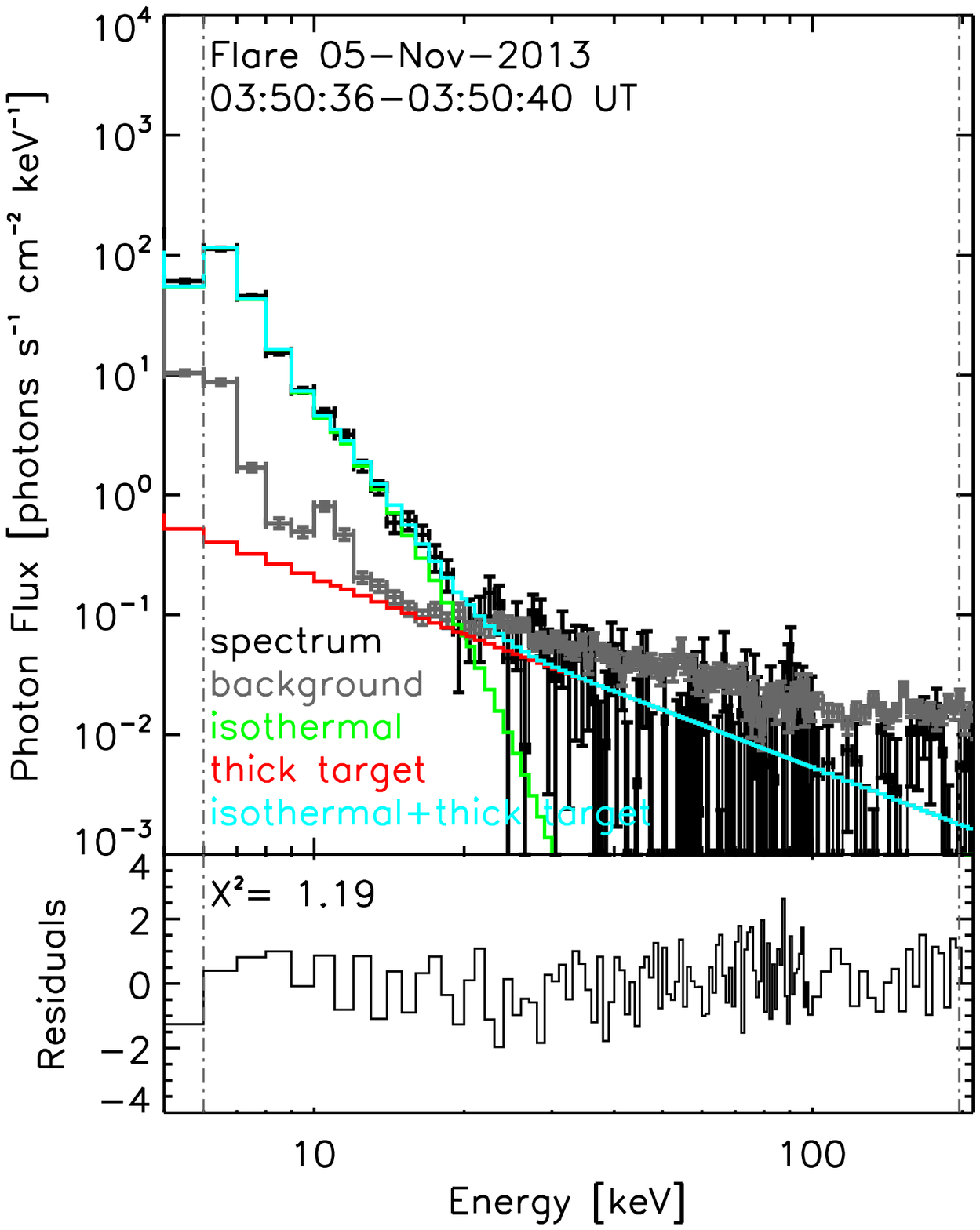}
\caption{Examples of the \rhessi\ fits (light blue histogram) with isothermal (green histogram) plus thick target model (red histogram) during the nonthermal peak at 03:50:20-03:50:24~UT(left panel) and a few seconds later at 03:50:36-03:50:40~UT (right panel). The \rhessi\ data and the background level are shown with black and gray histograms respectively. Bottom panels indicate residuals.
\label{Fig_RHESSI_fits}
}
\end{figure}

\begin{figure}\centering
\includegraphics[width=8cm]{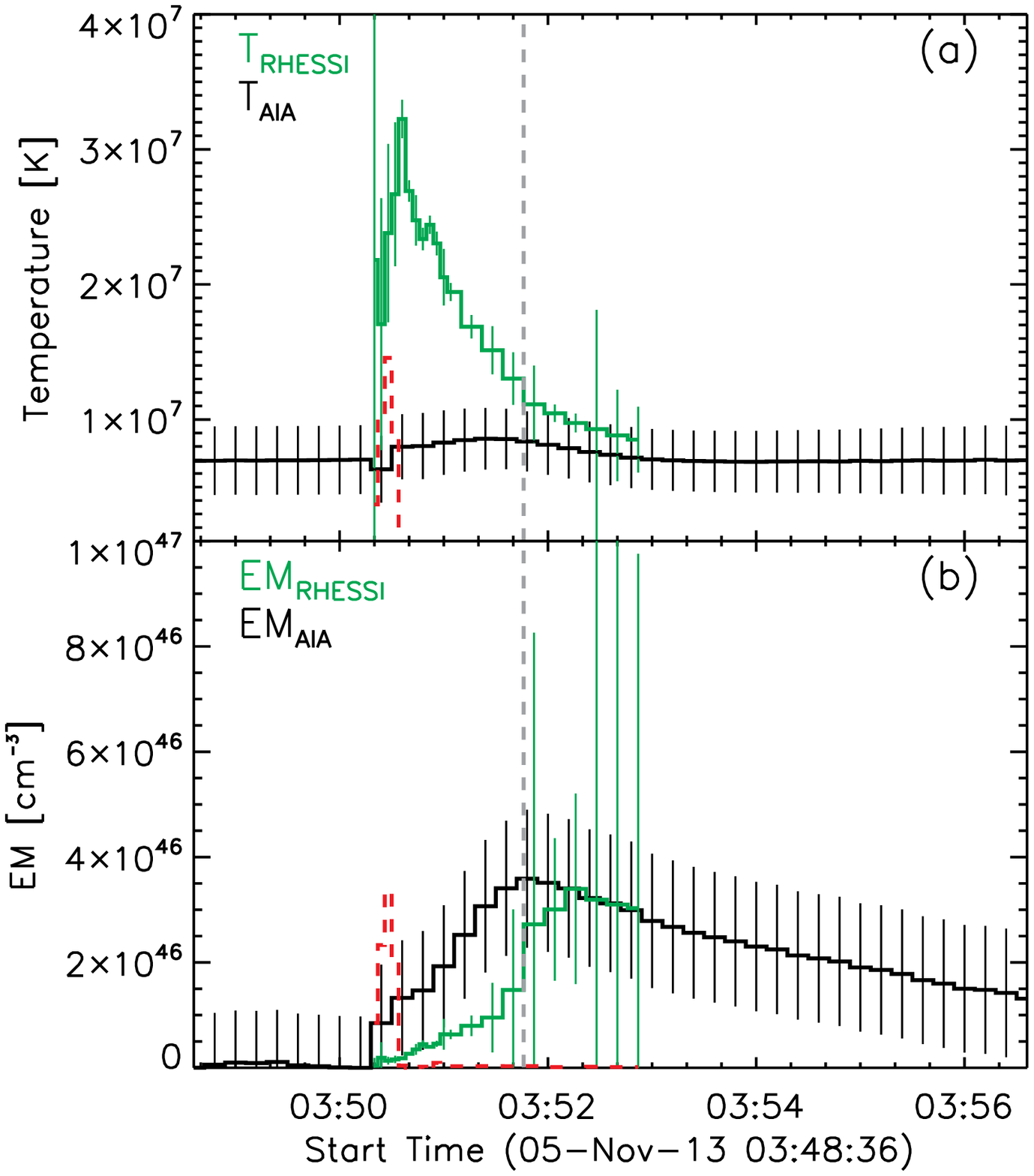}
\includegraphics[width=8cm]{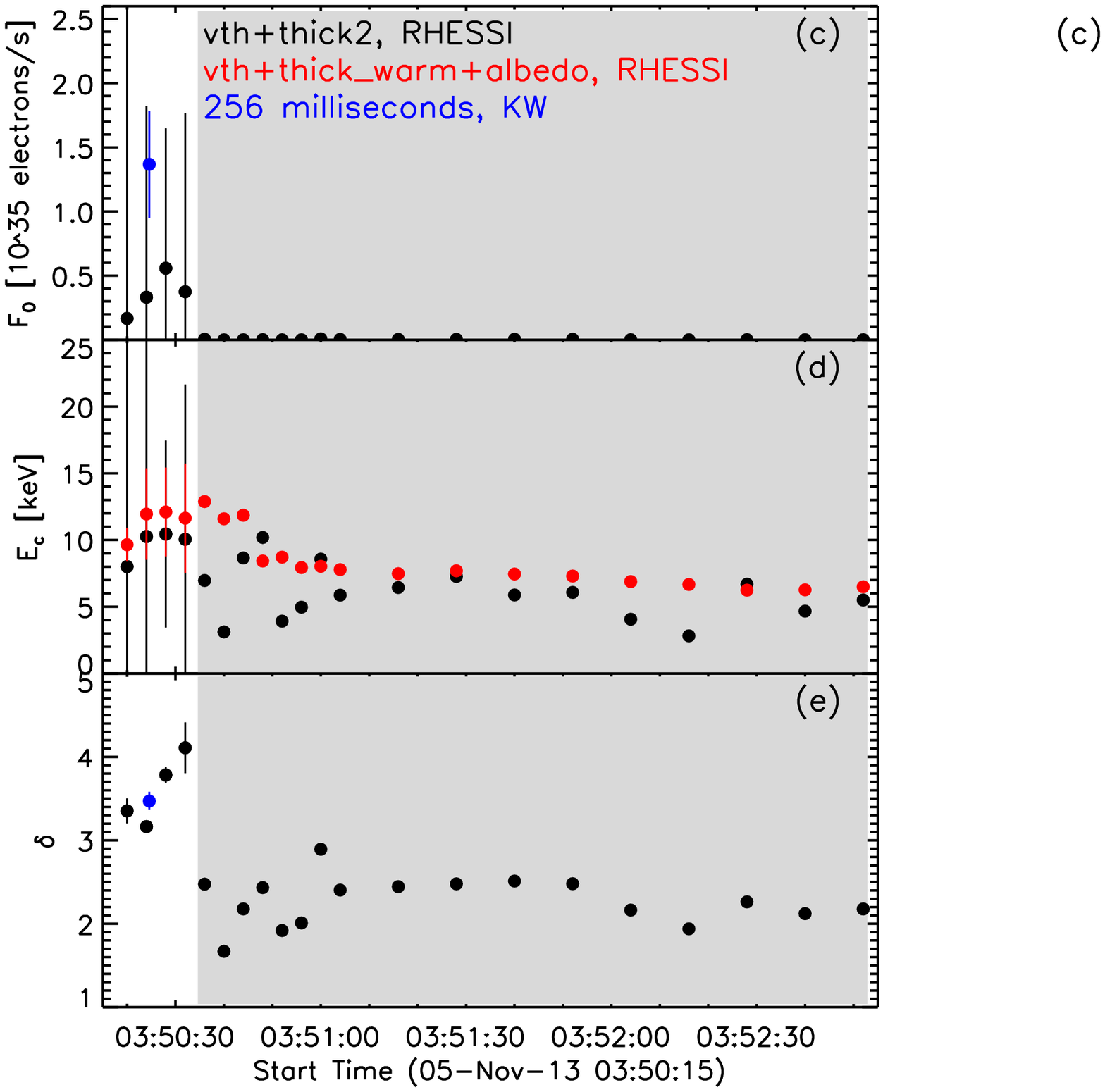}
\caption{Time evolution of $T$ {(a)}, $EM$ {(b), and nonthermal parameters (c, d, e)} for the November 5, 2013 flare. {(a)} \rhessi\ temperature (green histogram) from the isothermal model for spatially unresolved full solar disk and AIA temperature (black histogram) from the FOV shown in Fig.\ref{Fig_AIA_map}.  {(b)} \rhessi\ (green histogram) and preflare-subtracted AIA (black histogram) emission measure, inferred as described for the upper panel.
Vertical dashed line indicates the time of 03:51:46~UT, when $EM$ and $T$ from the \rhessi\ and the AIA data converge.
Red dashed histogram indicates the change of \rhessi\ nonthermal energy with time $dW^{\rm{RHESSI}}_{\rm{nonth}}/dt$ [arb. units] {(see Eq.\ref{eq5})}.
{(c) Total integrated electron flux $F_0$, (d) low-energy cut-off $E_c$, and
(e) spectral index $\delta$ obtained from the \rhessi\ {isothermal+thick target (black circles), isothermal+warm target+albedo (red circles),} and the \kw\ fits (blue circles).} Vertical lines indicate the range of 1$\sigma$ error on the fits of the \rhessi\ {and the \kw} data. The greyed out area indicates the time range at which no nonthermal component could be confidently identified in the data.
\label{Fig_EM_T}
}
\end{figure}

\subsubsection{Diagnostics using the regularized DEM maps}
\label{S_aia_diagn_maps}
The spatially resolved AIA data offer a much more detailed diagnostics of the thermal plasma than that performed above in Section~\ref{S_aia_diagn_integral}.
To this end, the same AIA data set (see section\,\ref{S_aia_diagn_integral}) was used to yield the spatial emission measure $EM^{\rm{AIA}}_{ij}$ and the mean temperature $\langle T^{\rm{AIA}}_{ij} \rangle$ in each pixel with coordinates $i=0,..., N_{\rm{px}}-1$ and $j=0,..., N_{\rm{px}}-1$, where the number of pixels $N_{\rm{px}}=51$ {in our case},
based on the regularized inversion code developed by \cite{2012A&A...539A.146H, 2013A&A...553A..10H}.
Because it attempts to solve an ill-posed problem, this algorithm can occasionally give alternating values, unless the DEMs have been forced to be positive. We checked that this does not typically happen within our flaring region, but mainly in some non-flaring pixels within FOV; thus, we made the inversion without forcing DEMs to be positive.
We found that negative values can occasionally appear at high temperatures, $T>25$~MK, to which AIA is almost insensitive.
For this reason, we restrict the range of the employed temperatures to 0.5--25~MK same as in Section\,\ref{S_aia_diagn_integral} (in contrast to the default 0.5--32~MK range).

Then, to subtract the non-flaring fore- and background plasma along each {line of sight} (LOS), we formed, for each pixel, the averaged distribution $\xi^{\rm{bk}}_{ij}(T_k)$ for nine first time intervals that correspond to the preflare phase (03:48:30-03:50:18~UT). This background DEM distribution $\xi^{\rm{bk}}_{ij}(T_k)$ was then subtracted from $\xi_{ij}(T_k)$ to form the DEM $\xi^{\rm{fl}}_{ij}(T_k)$, that pertains to the flare loops only. Note, that AIA has a gap of sensitivity to intermediate plasma temperatures about 5~MK; for that reason some of the background DEMs $\xi^{\rm{bk}}_{ij}(T_k)$ in this temperature range can happen to be larger than $\xi_{ij}(T_k)$, which might yield some negative $\xi^{\rm{fl}}_{ij}(T_k)$ values. To ensure that the flaring DEMs $\xi^{\rm{fl}}_{ij}(T_k)$ are positively defined, which is essential for further quantitative analysis, any obtained negative values were set to zero. This is justified by the fact that those negative values are always consistent with being zeros within their errors.
Now, the $EM^{\rm{AIA}}_{ij}$ and $\langle T^{\rm{AIA}}_{ij} \rangle$ maps for the flaring DEMs can be calculated similar to Eq.(\ref{eq1}) and Eq.(\ref{eq2})

\begin{equation}\label{eq3}
EM^{\rm{AIA}}_{ij}=S_{\rm{px}}\int\limits_{T_{\min}}^{T_{\max}} \xi^{\rm{fl}}_{ij}(T) dT,
\end{equation}

\begin{equation}\label{eq4}
\langle T^{\rm{AIA}}_{ij} \rangle =\frac{\int\limits_{T_{\min}}^{T_{\max}} T \xi^{\rm{fl}}_{ij}(T) dT}{\int\limits_{T_{\min}}^{T_{\max}} \xi^{\rm{fl}}_{ij}(T) dT},
\end{equation}
where $S_{\rm{px}}$ [cm$^2$] is the area of the AIA pixel.
The thermal energy density $w^{\rm{AIA}}_{ij}$ inferred for the FOV of interest
(see Fig.\ref{Fig_AIA_map}) can then be defined as
\begin{equation}\label{eq4b}
w^{\rm{AIA}}_{ij}  =3 k_B \langle T^{\rm{AIA}}_{ij} \rangle \sqrt{EM^{\rm{AIA}}_{ij}/(S_{\rm{px}}\ l_{\rm{depth}})} \; [\rm{erg\ cm^{-3}}],
\end{equation}
where $l_{\rm{depth}}=d_{\rm{depth}}\times 0.6 \times 7.25\times 10^{7}$ [cm] is the adopted length along the  LOS taken equal to the mean loop width $d_{\rm{depth}}=d_{\rm{width}}=4$ [px] and $k_B$ is the Boltzmann constant; both electron and ion contributions are accounted by using the factor of 3. The value $d_{\rm{width}}$ is the estimation of the loop width, which we found from inspection of the energy density images shown in  Figure \ref{Fig_EM_map}.
Animated Figure\,\ref{Fig_EM_map} shows time evolution of the temperature $\langle T^{\rm{AIA}}_{ij} \rangle$ (top left panel), the emission measure $EM^{\rm{AIA}}_{ij}$ (top right panel), chi-square ($\chi^2<2$, bottom left panel), and the thermal energy density $w^{\rm{AIA}}_{ij}$  (bottom right panel) from the area $A$.
We inferred these parameters for all times during the flare, the reference image in Figure~\ref{Fig_EM_map} shows the 13th time interval (03:51:06-03:51:18~UT). It is clearly seen that the mean temperature concentrates around 10 MK during the flare, while the emission measure and the thermal energy density change dramatically and have the peak right after the nonthermal HXR peak time with a subsequent decrease after it, prominently showing the Neupert effect.
\begin{figure}\centering
\includegraphics[width=0.95\columnwidth]{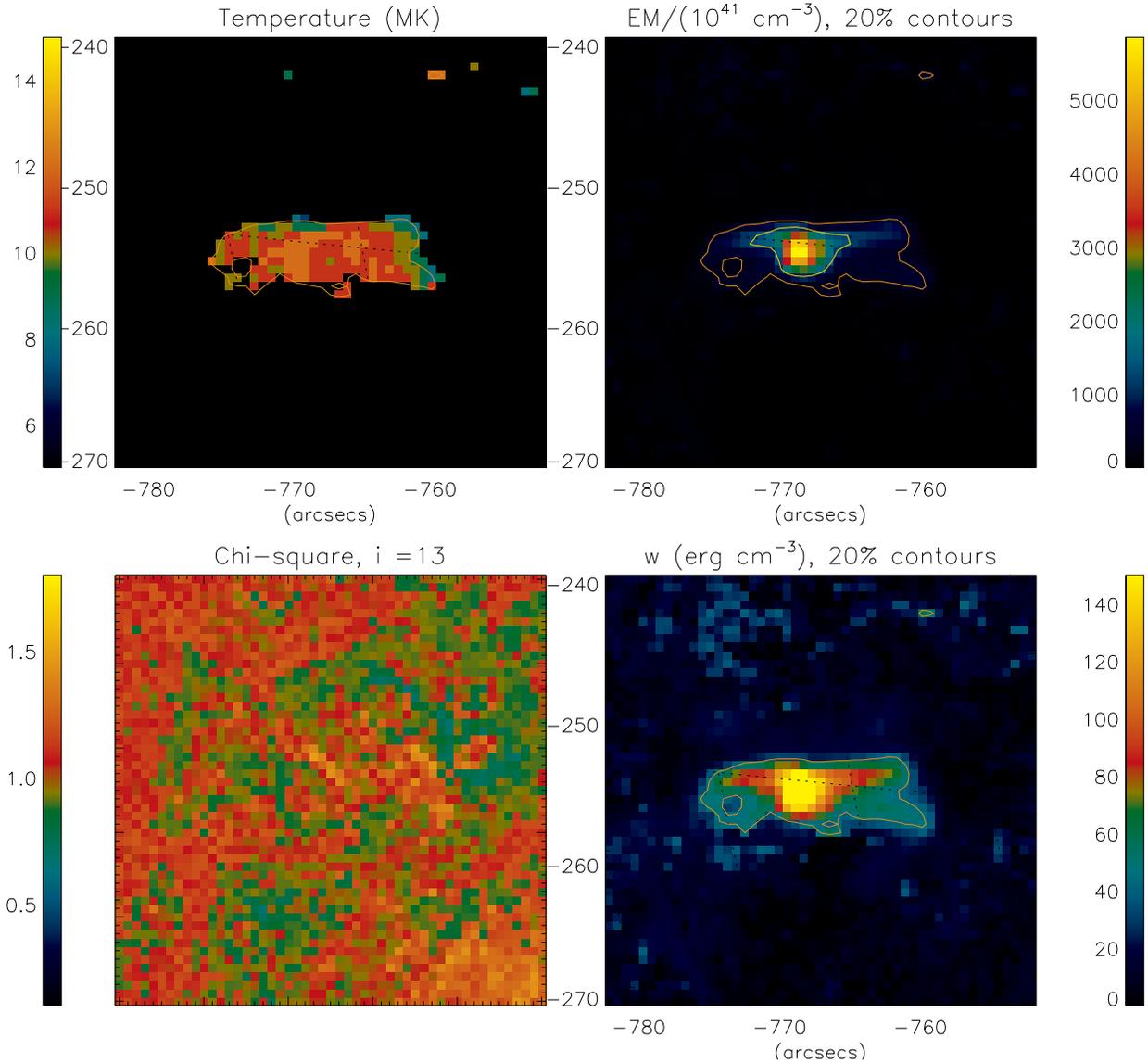}
\caption{Temporal evolution of plasma parameters obtained with {the regularized DEM maps} derived from \sdo/AIA data: the mean temperature map (top left panel), the emission measure map (top right panel), chi-square map (bottom left panel), and the thermal energy density map (bottom right panel) for the 13th time interval (03:51:06-03:51:18~UT). The orange contour indicates 20\% of the thermal energy density peak. Yellow contour indicates 20\%  of the emission measure peak. The mean temperature map is plotted only for pixels where the emission measure is greater than 5\%  of the maximum value of the emission measure for each time interval. On-line animation is available.
\label{Fig_EM_map}
}
\end{figure}

\subsection{Energy budget}
\label{S_energy_budget}
Now we have all inputs needed to quantify the thermal-to-nonthermal energy budget in this flare as complete as possible with the available data.
The nonthermal energy $W^{\rm{RHESSI}}_{\rm{nonth}}$ deposited to the flaring volume during the impulsive phase vs time  was computed as a cumulative sum using the parameters from the \rhessi\ fits:
\begin{equation}\label{eq5}
W^{\rm{RHESSI}}_{\rm{nonth}}=\int\limits_{-\infty}^t F_0 E_c \frac{\delta-1}{\delta-2} dt \; [\rm{erg}],
\end{equation}
where $F_0$, $E_c$, and $\delta$ are the thick target parameters (see Section\,\ref{S_rhessi_diagn}) from the fitting of the \rhessi\ data displayed in Figure\,\ref{Fig_EM_T}c-e;  $W^{\rm{RHESSI}}_{\rm{nonth}}$ is shown in blue in Figure \ref{Fig_Energy}.

The associated thermal energy can be computed from both the \rhessi\ and AIA data. To calculate the thermal energy detected by \rhessi, we employ the emission measure and temperature obtained from the \rhessi\ fit (see Section~\ref{S_rhessi_diagn} and Fig.~\ref{Fig_EM_T}):

\begin{equation}\label{eq6}
W^{\rm{RHESSI}}_{\rm{therm}}=3 k_B T_{\rm{RHESSI}} \sqrt{EM_{\rm{RHESSI}}\times V} \; [\rm{erg}],
\end{equation}
where $V$ is the volume of the corresponding thermal source, which is unavailable given that the source is only barely resolved by \rhessi.
Here we assume that the volume of the \rhessi\ thermal source is comparable to that of the AIA thermal source, which can be evaluated much more precisely. We then justify this assumption \textit{a posteriori} using 3D modeling (Section~\ref{S_modeling}).

The volume of the AIA thermal source can be estimated as
\begin{equation}\label{eq7}
V=S_{\rm{eff}}\times l_{\rm{depth}} \; [\rm{cm^3}],
\end{equation}
where the effective area is defined using the map of the  thermal energy density $w^{\rm{AIA}}_{ij}$ determined in Section~\ref{S_aia_diagn_maps} as:
\begin{equation}\label{eq8}
S_{\rm{eff}}=\frac{\left( S_{\rm{px}}\sum_{i=1}^{N_{\rm{px}}} \sum_{j=1}^{N_{\rm{px}}} w^{\rm{AIA}}_{ij} \right)^2}{ S_{\rm{px}}\sum_{i=1}^{N_{\rm{px}}} \sum_{j=1}^{N_{\rm{px}}} (w^{\rm{AIA}}_{ij})^2} \; [\rm{cm^2}].
\end{equation}
This estimate of the volume $V$ is only applicable during the course of the flare; not for the preflare phase. The thermal energy evolution $W^{\rm{RHESSI}}_{\rm{therm}}$ derived from the parameters of the \rhessi\ data fits using Eq.(\ref{eq6}) is shown in green in Figure \ref{Fig_Energy}.

The total thermal energy in the FOV derived from the AIA data is  computed from the spatial distribution of the thermal energy density $w^{\rm{AIA}}_{ij}$, obtained from the regularized DEM maps, by adding up the contributions from all pixels in the FOV:
\begin{equation}\label{eq9}
W^{\rm{AIA}}_{\rm{therm}}=S_{\rm{px}}\ l_{\rm{depth}}\sum_{i=1}^{N_{\rm{px}}} \sum_{j=1}^{N_{\rm{px}}} w^{\rm{AIA}}_{ij} \; [\rm{erg}],
\end{equation}
and then subtracting the minimal preflare value {$W^{\rm{AIA}}_{\rm{therm, min}}=6.35\times 10^{27}$}
\,[erg], which comes from the non-flaring pixels in the FOV.
Evolution of the AIA-derived thermal energy $W^{\rm{AIA}}_{\rm{therm}}$ is shown in black in Figure \ref{Fig_Energy}.

In agreement with our conclusions made in Section~\ref{S_aia_diagn_integral} (see Fig.~\ref{Fig_EM_T}), the thermal energy evolution could be subdivided into two stages.  At the first stage lasting roughly up to $\sim$03:51:46~UT, \rhessi\ and AIA see two distinct thermal sources: a hot and tenuous one, seen by \rhessi, and a cooler and denser one, seen by AIA. Interestingly, that initially the nonthermal energy (a blue histogram in Figure \ref{Fig_Energy}) is divided in roughly comparable portions between these two thermal sources. The sum of these two thermal energies roughly matches the available nonthermal energy deposition; thus the revealed evolutions of various energy components is quantitatively consistent with the thermal energy fully driven by the nonthermal energy deposition. No extra energy is needed to drive the observed plasma heating, while there is no unaccounted nonthermal energy to drive any other form of the energy, for example, the kinetic energy. Given that we do not see any eruptive activity associated with this flare, we have to conclude that, in this flare, the free magnetic energy was primarily converted to the nonthermal energy (acceleration of electrons), which, in its turn, was fully responsible for generation of the thermal energy component.

The hotter source seen by \rhessi\ displays a quicker cooling than the other thermal source (Fig.\,\ref{Fig_EM_T}a). Thus, at the second stage, after $\sim$03:51:46~UT, \rhessi\ started seeing the cooler and denser source--the same as AIA. The mismatch between the green and black curves in Figure~\ref{Fig_Energy} after 03:51:46~UT is about a factor of 2,
which can be considered as a measure of uncertainty in defining the thermal energy by two different instruments/techniques.
Formally, due to its large statistical errors, the \rhessi-derived thermal energy is not significantly different from that derived from the AIA data after 03:51:46~UT, even though the mean values are different by a factor of 2. In addition to the statistical uncertainties, there are also noticeable systematic uncertainties, particularly, because of poorly constrained volumes of the flaring loops. These uncertainties are difficult to firmly quantify; a reasonable ballpark estimate could be taken as a factor of 2, similar to the mismatch between the green and black curves after 03:51:46~UT.

\begin{figure}\centering
\includegraphics[width=0.7\columnwidth]
{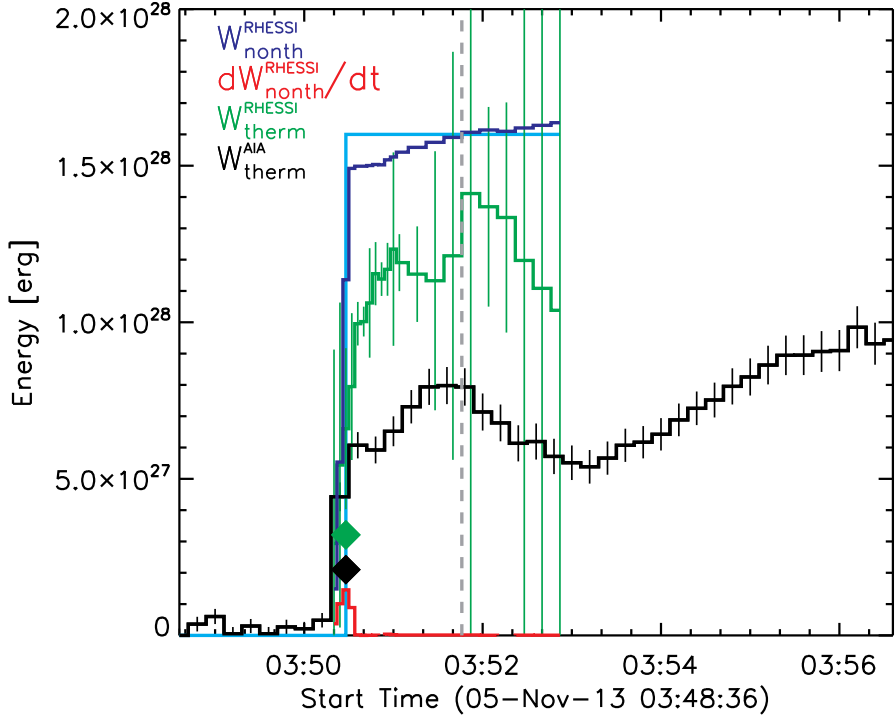}
\caption{ {Evolution of energy components in the November 5, 2013 flare. Cumulative nonthermal energy deposition  $W^{\rm{RHESSI}}_{\rm{nonth}}$ obtained using parameters of the nonthermal part of the \rhessi\ fits (Eq.\ref{eq5}) is shown in blue.
Thermal energy $W^{\rm{RHESSI}}_{\rm{therm}}$ (Eq.\ref{eq6}) computed using thermal part of the \rhessi\ fits is shown in green, while the thermal energy  $W^{\rm{AIA}}_{\rm{therm}}$ (Eq.\ref{eq9}) computed from the AIA DEM maps is shown in black. The red histogram shows the rate of \rhessi\ nonthermal energy deposition $dW^{\rm{RHESSI}}_{\rm{nonth}}/dt$ [arb. units]. Vertical dashed line indicates the time 03:51:46~UT the same as in Fig.\,\ref{Fig_EM_T}a,b.}
The green and black symbols indicate the model values of thermal energies for Loops 1 and 2, respectively, for 03:50:28~UT; see Table\,\ref{table_model_summary}. The light blue line shows the modeled total deposition of the nonthermal energy $\widetilde{W}$. For simplicity, the entire energy deposition is assigned here to the peak time of the impulsive phase.
\label{Fig_Energy}
}
\end{figure}

\section{Modelling of the November 5, 2013 flare}
\label{S_modeling}

To cross-check and validate our conclusions formulated above based mainly on analysis of EUV and X-ray data, {and, in particular, to estimate how accurate our selection of the volume of the hot flaring loop seen by \rhessi\ was,} we perform a 3D modelling and employ additional data: photospheric magnetograms,  \mw\ images and spectra. The 3D magnetic field model of the active region of interest is obtained using the automated model production pipeline \citep{2018ApJ...853...66N} based on the NLFFF extrapolation code \citep{2017ApJ...839...30F} initiated with \sdo/HMI  \citep{2012SoPh..275..207S} vector magnetogram (taken at 03:46:13.4~UT) that is used as the bottom boundary condition.  Using the IDL-based simulation tool, GX (Gyrosynchrotron / X-ray) simulator \citep{2015ApJ...799..236N, 2018ApJ...853...66N} we import the magnetic data cube, inspect the magnetic connectivity, and select two flux tubes roughly matching the available imaging data. These flux tubes are populated with thermal plasma and nonthermal electrons such as to match the available data for
a single time interval (03:50:28~UT) near the HXR peak as closely as possible \citep[see an example of such a modelling performed by][and references therein]{2018ApJ...852...32K}.

Once a model has been set up, GX Simulator permits computing
microwave, X-ray, and EUV emission
and compare them with the observations. {Recently, an ability to compute the thick-target X-ray emission has been added to GX Simulator, which now permits a much more realistic modeling of X-ray emission than was possible before.} The mismatches between the simulated and observed data (both images and spectra) are removed by adjusting the adopted model spatial distributions of the thermal and nonthermal particles and other their parameters such as density, temperature, and spectrum. Once the simulated emissions match the observations closely, the model is deemed validated.

Figure~\ref{Fig_model_0} shows the final validated model.
{Although both thermal plasma and nonthermal electrons are present in both model flaring loops, these loops make dissimilar contributions to emission in different domains. Flux tube 1 (Figure~\ref{Fig_model_0}) is hotter and more tenuous, so it dominates X-ray spectrum, though flux tube 2 (Figure~\ref{Fig_model_0}) makes a non-negligible contribution to the X-ray emission. Flux tube 2 is cooler and denser; it dominates EUV emission detected by AIA. Both flux tubes contain comparable numbers of nonthermal electrons; however, given that the energy spectrum of the nonthermal electrons in flux tube 2 is noticeably harder than in flux tube 1, flux tube 2 makes a dominant contribution to the microwave GS emission. }
Individual contributions of our two flaring loops to the X-ray and \mw\ spectra are shown in Figure~\ref{Fig_model_radio_spec} by the dashed and dotted lines.

Figure~\ref{Fig_model_radio_spec} shows, by the solid curves, a perfect match between simulated and observational spectral data in both X-ray and \mw\ domains. Comparison between the simulated and observed images requires additional co-alignment to make up for simulated image position errors. These errors originate from two sources. The first of them is a geometrical error due to projection effects (which is getting larger towards the limb) and also to distortions because mismatch between the spherical shape of the Sun and Cartesian coordinate system adopted for the modelling. The second source of error is an uncertainty in reproducing the magnetic field lines within the NLFFF data cube: on average the spatial deviation between the reconstructed and ``true'' field line has been estimated to be about 10\% of the loop length \citep{2019ApJ...870..101F}.

In Section~\ref{S_imaging} we
co-aligned the AIA images to match the \rhessi\ ones by adjusting the roll angle. Accordingly, here we co-align simulated X-ray images to match the observed \rhessi\ images by applying a shift of
$dx=-1.5\arcsec$, $dy=-8.6\arcsec$ (Figure \ref{Fig_xray_image}); the same shift was then applied to the simulated \mw\ images for consistency.
This shift is in the range of the expected model positioning error described above.

\Mw\ images are obtained with NoRH, which itself might have a positioning error within $10\arcsec$ over both $x$ and $y$ coordinates; thus, the \mw\ images (see Figure \ref{Fig_radio_image}) were further co-aligned with the (shifted) model images by shifting the observed ones by $dx=-3.5\arcsec$, $dy=-5.7\arcsec$,  at 17\,GHz, and by $dx=-5\arcsec$, $dy=2\arcsec$, at 34\,GHz, which is within the NoRH positioning accuracy.

These Figures show a remarkable agreement between the simulated and observed spectra and images in both \mw\ and X-ray domains, and so validate the entire 3D model.
This implies that physical picture developed based on only a subset of available data, namely, EUV and X-ray, is also consistent with other context data---photospheric magnetic field measurements and \mw\ imaging and spectroscopy; thus, it captures at least the main flare properties and so can be conclusively used to quantify this flare in 3D.

{Using the model, validated by comparison between simulated and observed emissions, we cross-check our assumptions and findings derived from the data analysis. Most relevant numbers of the model are summarized in Table~\ref{table_model_summary}. In Section~\ref{S_rhessi_diagn}, we adopted that both flux tubes have the same volume.
In the model, the flux tube volumes are indeed comparable, although not identical: they differ almost by a factor of 3, which validates our assumptions within the same factor for the volume itself and within a factor of $\sim\sqrt{3}$ for the energy.

Let us now evaluate if the amount of the nonthermal energy in the model is sufficient to supply the observed thermal energy in both loops. The model contains instantly $\sim4\times10^{26}$\,erg of the total nonthermal energy, which is divided roughly equally between these two loops. To estimate the total deposition of the nonthermal energy $\widetilde{W}$, we have to estimate escape time $\tau$ of the nonthermal electrons from the loops and the total duration of the nonthermal energy injection $\Delta t$ (at the level of half maximum): $\widetilde{W}\approx \Delta t W_{nth} / \tau$. To estimate the escape time, we adopt that the nonthermal electrons escape the loops more or less freely, such as $\tau \sim l/v\sim 0.2-0.3$\,s, where $l\approx1.5\cdot10^9$\,cm (see Table~\ref{table_model_summary}) for any of the loops and $v\approx0.2c$ is the velocity of 10\,keV electrons. From the full duration of the nonthermal X-ray and radio emission, $\approx15$\,s, we find $\Delta t\approx 8$\,s. Putting all numbers together, we obtain $\widetilde{W}\approx 1.6\times10^{28}$\,erg (Figure~\ref{Fig_Energy}, light blue line) in a perfect agreement with the nonthermal energy deposition derived from the X-ray data analysis (Figure~\ref{Fig_Energy}, dark blue line). The thermal energies of both flaring loops are obtained by direct integration over the model volume within the GX Simulator tool. They are shown by green and black symbols in Figure\,\ref{Fig_Energy} and match well those derived from the data. Thus, our data-constrained and data-validated 3D model agrees well with the results of the data analysis performed in Section\,\ref{S_Observations}.
}

\begin{table}[ht]
\caption{Summary of the 3D model}
\begin{tabular}{l l l l}
\hline\hline
Parameter & Symbol, units &  Loop I &  Loop II\\ [0.5ex]
\hline
{\textit{Geometry}:} &  & \\
\quad Length of the Central Field Line     & $l$, cm  & $1.46\times10^9$ & $1.62\times10^9$  \\
\quad Model Volume; $\left[\int n_0 dV\right]^2/\int n_0^2 dV$ & $V$, cm$^3$ & $6.86\times10^{25}$ & $2.39\times10^{25}$  \\
{\textit{Thermal plasma}:} &  & \\
\quad Emission Measure, $\int n_0^2 dV$ & $EM$,  cm$^{-3}$ & $1.40\times10^{45}$ & $1.69\times10^{46}$  \\
\quad Mean Number Density, $\int n_0^2 dV/\int n_0 dV$ & $n_{th}$, cm$^{-3}$ & $0.45\times10^{10}$ & $2.66\times10^{10}$ \\
\quad Temperature      & $T$, MK  &  25 & 8  \\
\quad Instant Total Thermal Energy & $W_{th}$, erg &  $3.21\times10^{27}$ & $2.1\times10^{27}$ \\
\textit{Nonthermal electrons:} &  &  \\
\quad Total electron number & $N_b$, cm$^{-3}$ &  $8.1\times10^{33}$ & $7\times10^{33}$   \\
\quad Low-energy Cutoff & $E_0$, keV & 10  & 10   \\
\quad Spectral Index & $\delta$ & 4.3 & 3.3 \\
\quad Instant Total Nonthermal Energy & $W_{nth}$, erg &  $1.85\times10^{26}$ & $2\times10^{26}$ \\
[1ex]
\hline

\end{tabular}

\label{table_model_summary}
\end{table}


\begin{figure}\centering
\includegraphics[width=10cm]{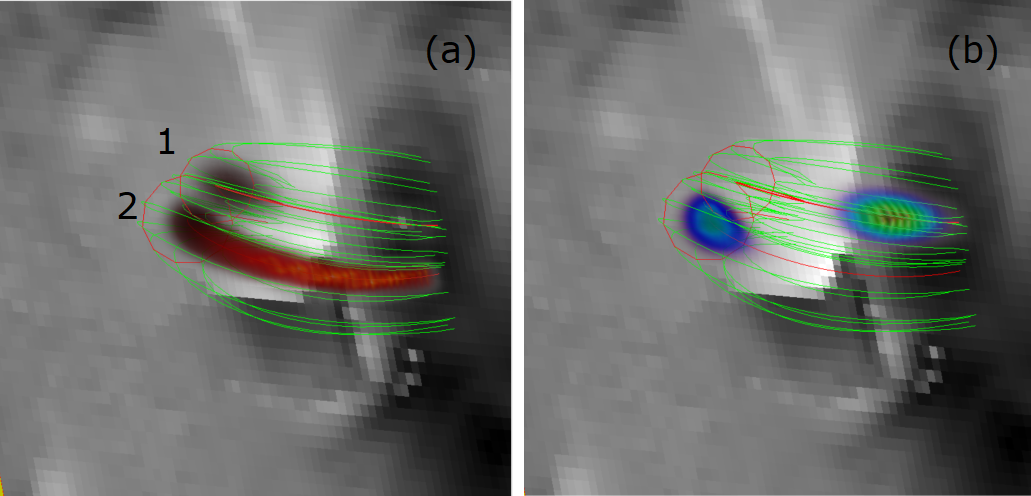}
\caption{The 3D model with two flux tubes (closed loops) filled with thermal plasma {(a)} and nonthermal electrons {(b) for the HXR impulse time (03:50:28~UT) overlaid on the LOS HMI magnetogram. Green lines visualize the closed field lines, the red lines and circles indicate corresponding central (reference) field lines and the cross-sectional areas of the flux tubes respectively. (a) The spatial distributions of thermal plasma in the flux tubes is shown by the red volume. The digits indicate the flux tube number. (b) The spatial distribution of nonthermal electrons in two flux tubes is shown by the blue-green volume.}
\label{Fig_model_0}
}
\end{figure}

\begin{figure}\centering
\includegraphics[width=0.47\columnwidth]{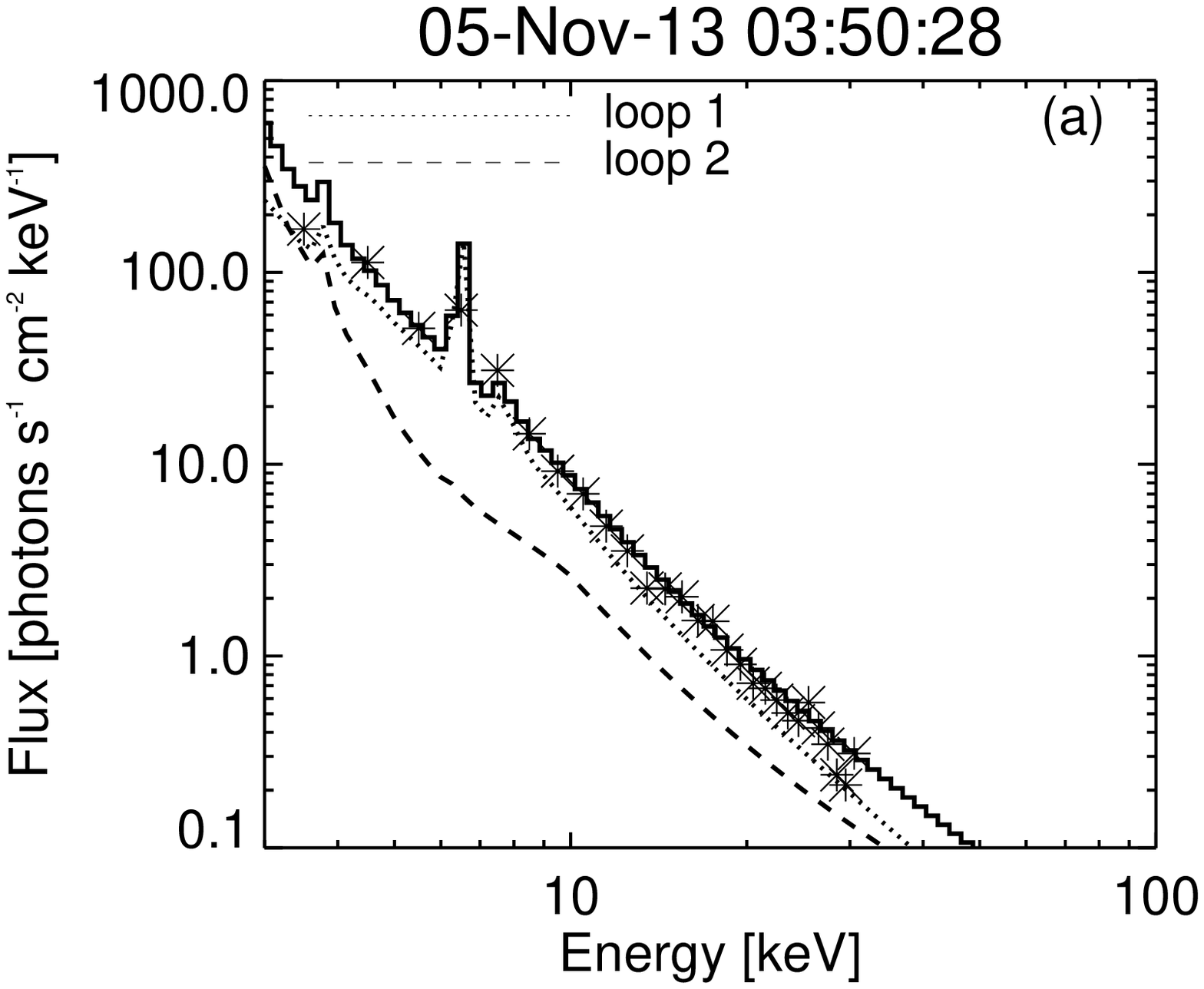}
\includegraphics[width=0.47\columnwidth]{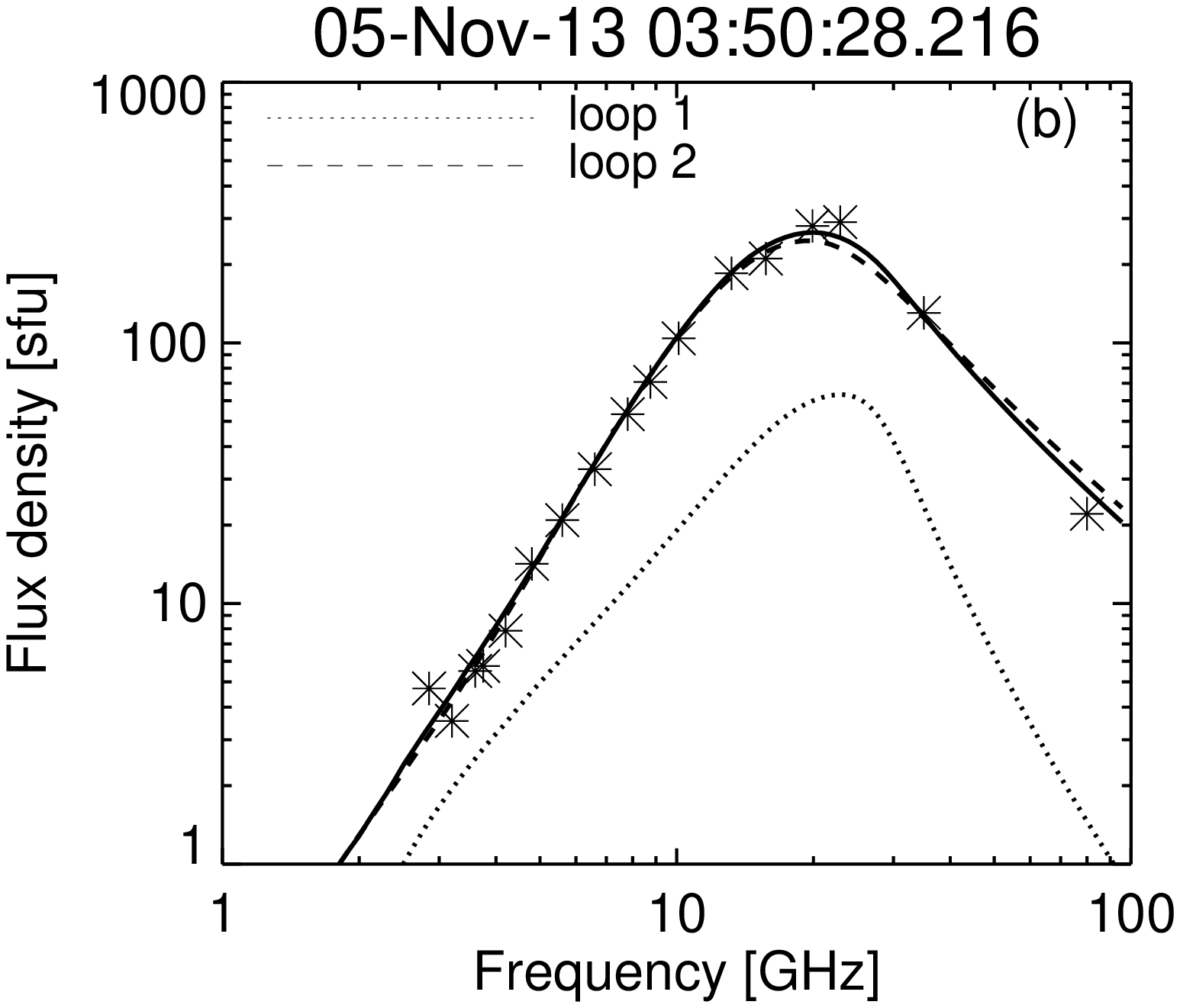}
\caption{(a) The observed {\rhessi} X-ray spectrum for time interval 03:50:28-03:50:32 UT (asterisks) and the corresponding simulated spectrum (histogram) from the 3D model; (b) the same for the radio domain with NoRP and SRS data at 03:50:28 UT (asterisks) and the simulated microwave spectrum  (black solid line). {The dotted and dashed lines correspond to contribution from loop 1 and loop 2 separately to the X-ray and the radio simulated spectra.}
\label{Fig_model_radio_spec}
}
\end{figure}

\begin{figure}\centering
\includegraphics[width=0.46\columnwidth]{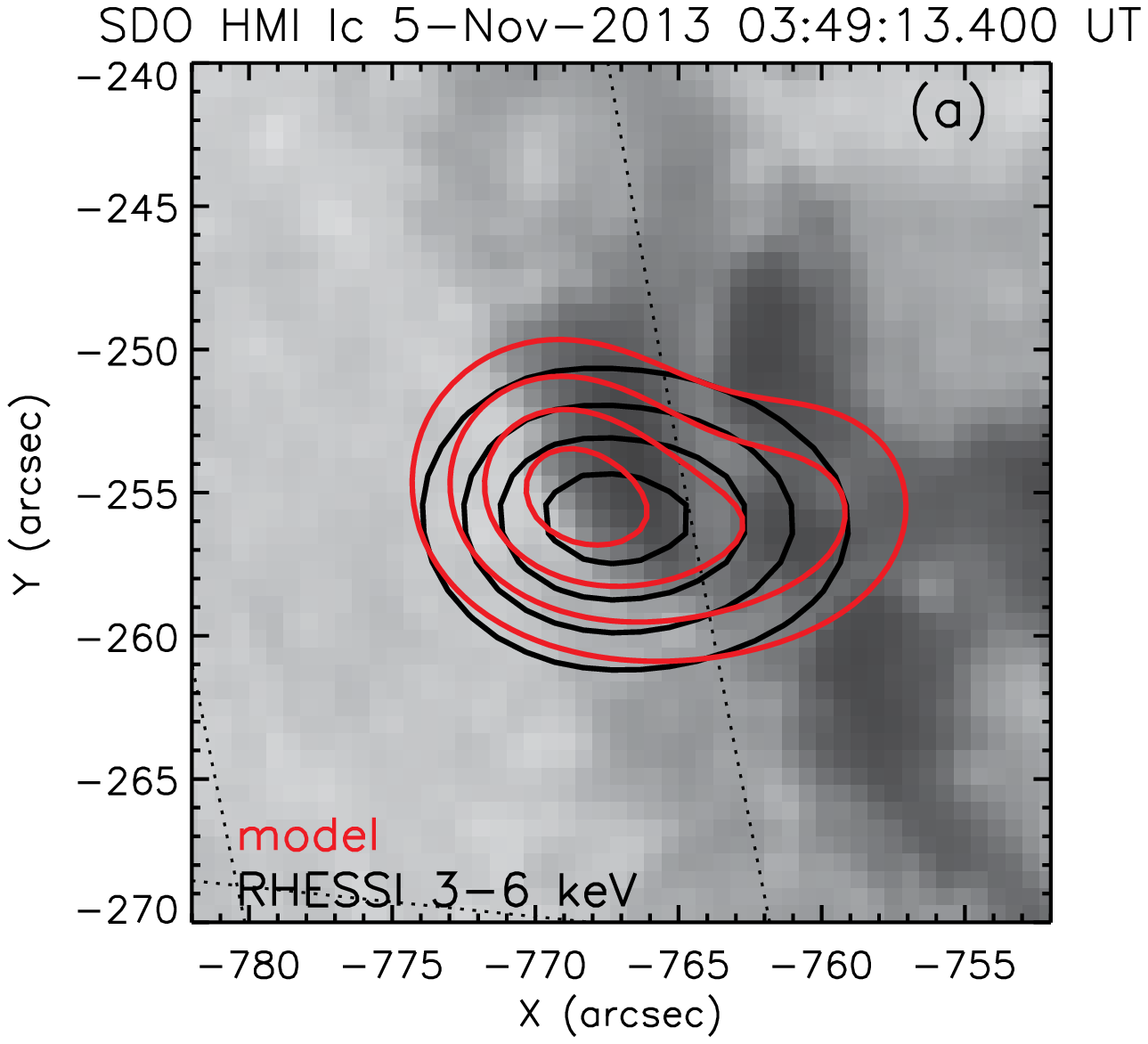}
\includegraphics[width=0.46\columnwidth]
{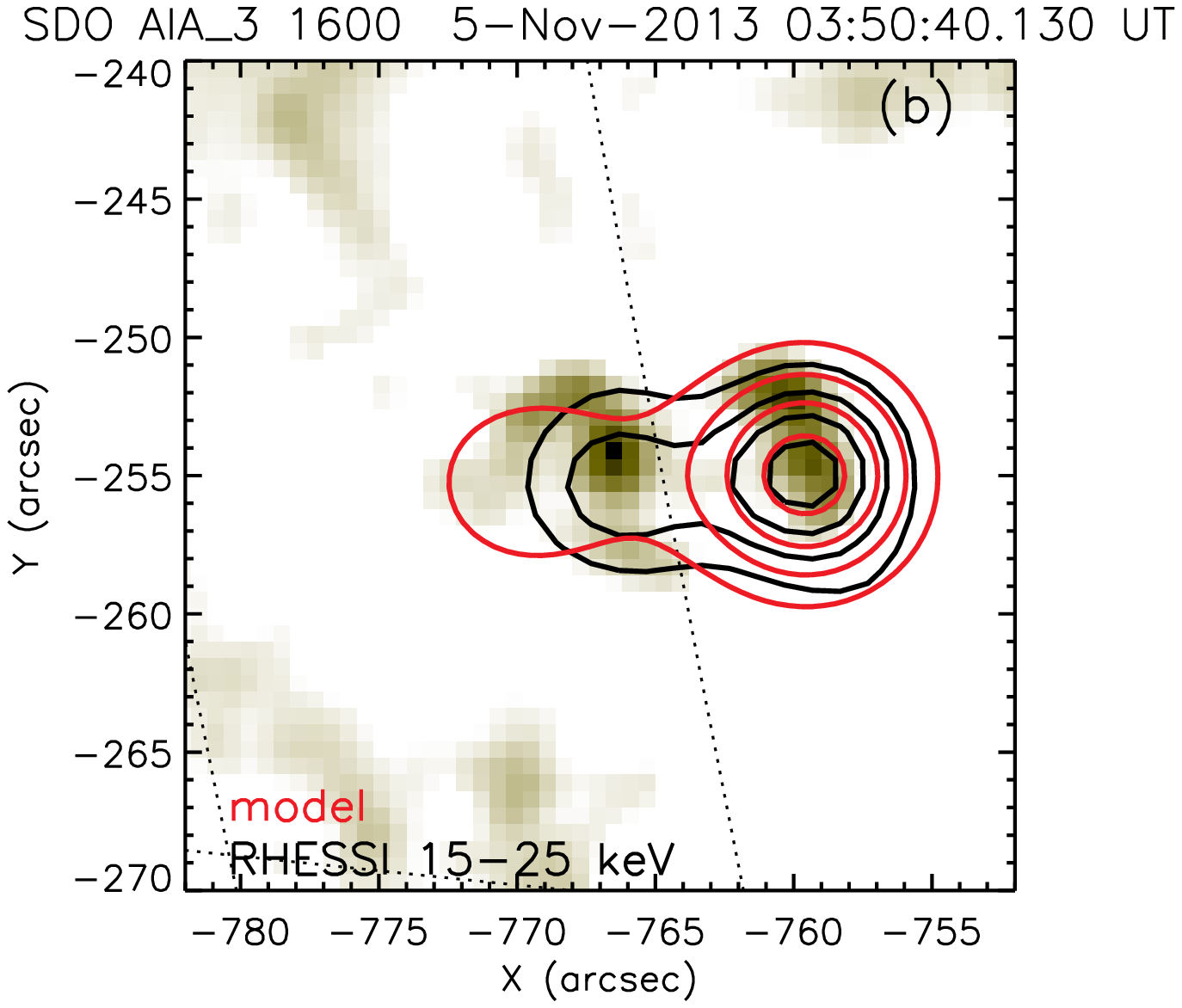}
\caption{The \rhessi\ CLEAN 3-6 keV {(a)} and 15-25 keV {(b)} 30, 50, 70, 90\% contours (black lines) and PSF-convolved model image contours (red lines) for the same energies for time interval 03:50:20-03:50:40~UT {overlaid on (a) an \sdo/HMI white light image and (b)
{AIA map in 1600 \AA\ wavelength channel.}
To co-align the \rhessi\ and HMI map the HMI roll angle has been changed similar to the AIA maps way HMI\_roll\_angle$=0.11$.
For co-alignment of the model X-ray and the \rhessi\ images a shift of $dx=-1.5\arcsec$, $dy=-8.6\arcsec$ has been applied to the simulated images as explained in the text.}
\label{Fig_xray_image}
}
\end{figure}

\begin{figure}\centering
\includegraphics[width=0.46\columnwidth]{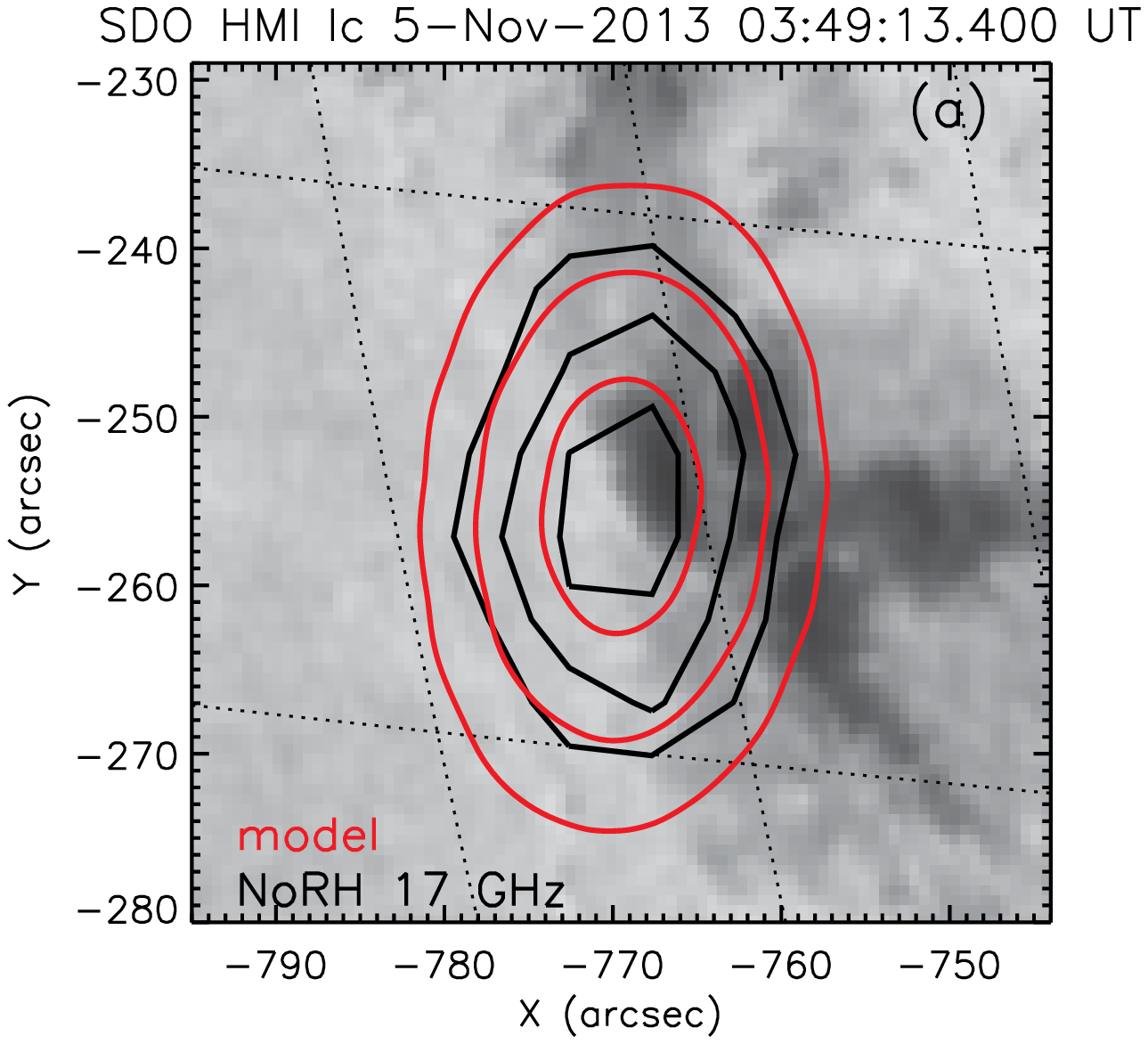}
\includegraphics[width=0.46\columnwidth]{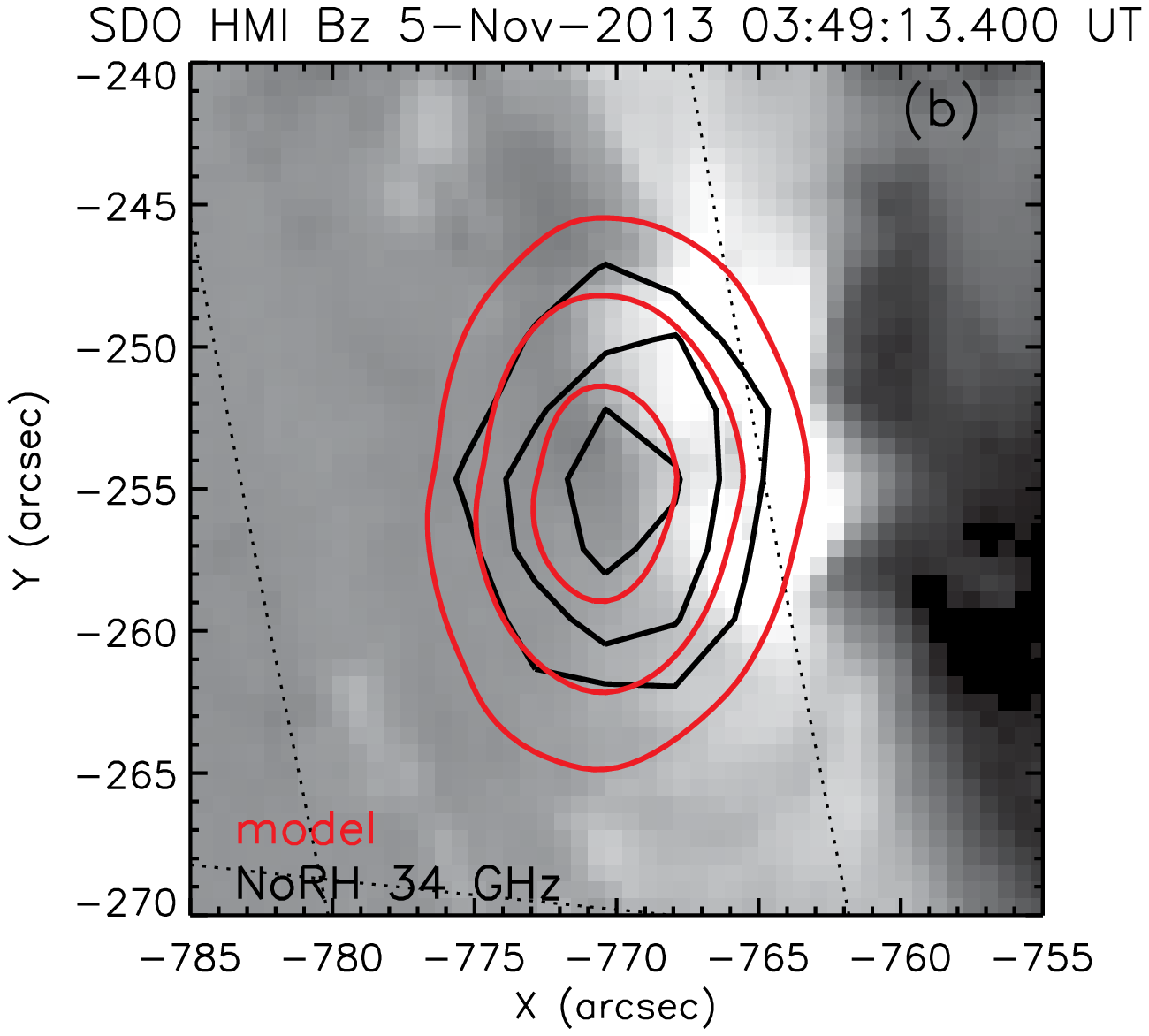}
\caption{The NoRH 17 GHz (a) and 34 GHz (b) 50, 70, and 90\% contours (black lines) and  PSF-convolved model image contours (red lines) at the same frequencies {overlaid on a \sdo/HMI white light image (a) and LOS magnetogram (b). To co-align the images we applied: (i) HMI\_roll\_angle$=0.11$ same as for AIA images; (ii) a shift of $dx=-1.5\arcsec$, $dy=-8.6\arcsec$ to the model images following Fig.~\ref{Fig_xray_image}; (iii) a shift of $dx=-3.5\arcsec$, $dy=-5.7\arcsec$  for  NoRH 17~GHz image and $dx=-5\arcsec$, $dy=2\arcsec$, for the 34~GHz  NoRH image.}
\label{Fig_radio_image}
}
\end{figure}

\section{Discussion and conclusions}
\label{S_disc_concl}

\citet{2005ApJ...621..482V} proposed to distinguish between the ``empirical Neupert effect'' (ENE)  vs  the ``theoretical Neupert effect'' (TNE), where ENE refers to relationships between the observed thermal and nonthermal \textit{emissions}, while TNE refers to the inferred thermal and nonthermal \textit{energies}. In contrast to their expectation, however, \citet{2005ApJ...621..482V} found that ENE shows a noticeably higher statistical significance than TNE. Two main sources of the lower TNE significance were proposed: (i) a potential presence of additional plasma heating unrelated to the nonthermal electron loss and (ii) uncertainty in quantification of the nonthermal energy related primarily to the uncertainty of low-energy cutoff in the spectrum of flare-accelerated electrons used to be derived from the X-ray thick-target spectral fit.
Therefore, to quantify the TNE much more conclusively requires that those two uncertainties have been removed or at least minimized.
This requirement is well fulfilled for the `cold' flares  \citep{2018ApJ...856..111L}.
Indeed, these flares show the weakest thermal response, compared with other flares, do not show any ``direct'' preflare heating, and the nonthermal X-ray emission dominates down to unusually low energies enabling to recover the low-energy cutoff values a factor of a few lower than in an average flare. In addition, the low-energy cut-off can now be much better constrained using a recently developed warm-target fit \citep{2019ApJ...871..225K}.

In this paper we have performed a detailed analysis of one `cold' flare, SOL2013-11-05T035054, which is ideal for such a case study because in addition to the mentioned advantages common for all flares from this subclass, it was observed with a combination of instruments sensitive to the nonthermal component as well as both hot and warm plasma. Furthermore, the flare displayed a simple single rise-and-fall time profiles, which simplifies analysis of the energy component evolution enormously compared with flares having  complex light curves.


 We have found that morphologically the flare consisted of two reasonably compact flux tubes, where a single episode of electron acceleration happened.
Thus, the flaring energy release is likely driven by inductive interaction resulting in magnetic reconnection between these two flux tubes. The accelerated nonthermal electrons have their energy roughly equally between these two flaring loops. However, the spectral slopes of these two nonthermal electron populations are different from each other: it is harder in cooler loop 2 than in hot loop 1 (see Table\,\ref{table_model_summary}). Although we cannot nail down the exact reason for that, we believe that some sort of the energy-dependent electron transport from the acceleration region to the loops must be involved, perhaps, due to wave-particle interactions.

It is further interesting that in spite of comparable amounts of the nonthermal energy deposited to both these loops (resulting in comparable amounts of the thermal energy), their specific thermal responses are rather dissimilar. Indeed, one loop remained tenuous with only a modest increase of $EM$ but became hot with temperature up to $\sim$30~MK (that seen by \rhessi); the other one displayed a larger $EM$, but was heated more gently (only up to $\sim10$~MK).
In the course of the flare, the hotter and more tenuous flux tube experienced a faster cooling (in agreement with the anticipated faster rate of the heat conduction); thus, the cooler flux tube survived longer and dominated the late phase of the post-flare cooling.

It is puzzling why these two flux tubes react that differently in response on comparable impacts from nonthermal electrons.
There could be several possible causes for that; however, it is difficult to understand within the chromospheric evaporation alone without considering nonthermal electron loss in the coronal portion of the flaring loops. Firstly, the spectral slopes of the nonthermal electrons are different in these two loops (see Table\,\ref{table_model_summary}). Simulations of the chromospheric response in flaring loops \citep{1984ApJ...279..896N, 1985ApJ...289..414F, 1989ApJ...341.1067M, 2015ApJ...808..177R}  do show a dependence of the response on the spectral slope: softer nonthermal spectra yield more efficient evaporation as they deposit their nonthermal energy at higher chromospheric heights. However, in our case, we see exactly the opposite trend: the loop with softer nonthermal spectrum remains more tenuous. Secondly, the energy deposited into the chromospheric footpoints depends on the flux tube geometry; in particular, on the footpoint area, which itself depends on the mirror ratio of the loop. Interestingly, within our 3D model, both flaring loops have comparable mirror ratios about 3, which does not favor one loop over the other in terms of nonthermal electron precipitation to the footpoints. Thirdly, a beam-like pitch-angle anisotropy, if present, may enhance the precipitation. Again, we do not see any evidence in favor of that: the footpoint HXR source in loop 1 is stronger than that in loop 2, which implies that the nonthermal electron precipitation was stronger in the tenuous loop (loop 1). This implies that the chromospheric evaporation took place in loop 1, while worked in a regime of only a modest enhancement of the coronal plasma density, but efficient heating. Most likely, loop 2 was denser from the beginning, so the lowest-energy nonthermal electrons lost their energy in the coronal portion of loop 2, thus providing a modest heating of this plasma, while the chromospheric evaporation was driven by less numerous higher-energy electrons, resulting in a more gentle evaporation. A combination of these two processes likely resulted in a cooler and denser plasma in loop 2 compared with loop 1.

It is remarkable that the nonthermal energy deposition ($\sim1.5\times10^{28}$~erg) over the impulsive flare phase matches nicely the sum of the thermal energies of the two flaring flux tubes ($\sim1\times10^{28}$~erg in the hotter one and $\sim0.5\times10^{28}$~erg in the cooler one). This implies, that within the uncertainties, the observed thermal energy is entirely supplied by the nonthermal electrons, i.e., displays a clear picture of TNE. There is no statistically significant room for other sources of energy (e.g., direct heating or acceleration of ions),
which would drive an extra heating and  evaporation.
Similarly, there is no indication of any other secondary energies (rather than the thermal one) such as kinetic energy of regular motion (e.g., an eruption).

\acknowledgements
This work was partly supported by 
by NSF grant AGS-1817277
and NASA grants
80NSSC18K0667,
80NSSC19K0068,
and 80NSSC18K1128
to New Jersey Institute of Technology.
GM was supported by the project RVO:67985815 and the project LM2015067: EU-ARC.CZ - National
Research Infrastructure by Ministry of Education of the Czech Republic.
EPK was supported by STFC consolidated grant  ST/P000533/1. {We are thankful to Alexandra Lysenko for fruitful discussions of the \kw\ data.}

\bibliography{all_issi_references,fleishman}

\appendix
\section{Statistical uncertainties of the involved parameters}\label{A_appendix_errors}

Both OSPEX and regularized inversion provide statistical errors of the output parameters. In this paper we, however, use a number of measures constructed from a combination of the immediate outcome of the codes, such as the energy densities, whose statistical uncertainties must be computed from the error propagation.

To compute the statistical uncertainty of the emission measure ($EM$) obtained from DEM distribution using Eq.\,(\ref{eq1}), we note that the DEMs there can be expressed as a sum of a mean $\overline{\xi_i (T_i)}$ and a fluctuating $\delta \xi_i (T_i)$ components: $\xi_i (T_i)=\overline{\xi_i (T_i)} +\delta \xi_i (T_i)$ and assume that the fluctuating components are not correlated between each other so that $\left<\delta \xi_i \delta \xi_j\right>=0$ for $i\neq j$. In this case we can compute the standard deviation of a sum $g=\sum_{i=1}^{N}g_i$ of uncorrelated variables $g_i$ from their standard deviations $\delta g_i$ as
\begin{equation}\label{eqa0b}
\delta g= \sqrt{\sum_{i=1}^{N} (\delta g_i  )^2 },
\end{equation}
which, being applied to  Eq.\,(\ref{eq1}) yields
\begin{equation}\label{eqa1}
\delta EM_{\rm{AIA}}= A\sqrt{\sum_i \overline{\delta \xi_i^2} (T_i) \Delta T_i^2},
\end{equation}
where $\overline{\delta \xi_i} (T_i)$ are
the DEM errors for  temperatures $T_i$ in the range $0.5-25$~MK, obtained from the regularization method by \citet{2012A&A...539A.146H, 2013A&A...553A..10H}.

The temperature $\langle T_{\rm{AIA}} \rangle$ defined by Eq.\,(\ref{eq2}) is a ratio of two values that both contain uncertainties. In this case it is convenient to use a general equation for the (independent) error propagation
\begin{equation}\label{eqa0a}
\delta f =
\sqrt{\left ( \frac{\partial f}{\partial x} \right )^2 \delta _x^2 + \left ( \frac{\partial f}{\partial y} \right )^2 \delta _y^2},
\end{equation}
where $\delta _x$ and $\delta _y$ represent the standard deviation of the random $x$ and $y$ values, respectively, which, being applied to Eq.\,(\ref{eq2}), yields:

\begin{equation}\label{eqa2b}
\delta \langle T_{\rm{AIA}} \rangle=\sqrt{\frac{A^2}{EM_{\rm{AIA}}^2}\sum_i T_i^2 \overline{\delta \xi_i^2} (T_i) \Delta T_i^2 + \frac{\langle T_{\rm{AIA}} \rangle ^2}{EM_{\rm{AIA}}^2} \delta EM_{\rm{AIA}}^2}.
\end{equation}
Similar equations are valid for the emission measure and the temperature defined for individual pixels from the DEM maps, Eqs.\,(\ref{eq3}, \ref{eq4}).

The statistical errors of the thermal energy density defined by Eq.(\ref{eq4b}) from the AIA data can also be computed following Eq.\,(\ref{eqa0a}) from the standard deviations of $EM^{\rm{AIA}}_{ij}$ and $\langle T^{\rm{AIA}}_{ij} \rangle$
\begin{equation}\label{eqa3a}
\delta w^{\rm{AIA}}_{ij}  = \sqrt{\left ( \frac{\partial w^{\rm{AIA}}_{ij}}{\partial \langle T^{\rm{AIA}}_{ij} \rangle} \right )^2
( \delta  \langle T^{\rm{AIA}}_{ij} \rangle)^2 +
\left ( \frac{\partial w^{\rm{AIA}}_{ij}}{\partial EM^{\rm{AIA}}_{ij}} \right )^2
( \delta  EM^{\rm{AIA}}_{ij})^2},
\end{equation}
where
$\partial w^{\rm{AIA}}_{ij} / \partial EM^{\rm{AIA}}_{ij} = 3 k_B   (S_{\rm{px}}\ l_{\rm{depth}})^{-0.5}  \langle T^{\rm{AIA}}_{ij} \rangle / (2 \sqrt{EM^{\rm{AIA}}_{ij}})$
and $\partial w^{\rm{AIA}}_{ij} / \partial \langle T^{\rm{AIA}}_{ij} \rangle =\\ 3 k_B   (S_{\rm{px}}\ l_{\rm{depth}})^{-0.5} \sqrt{EM^{\rm{AIA}}_{ij}}$. 
Substitution of these derivatives into Eq.\,(\ref{eqa3a}) yields:
\begin{equation}\label{eqa3}
\delta w^{\rm{AIA}}_{ij}  = 3 k_B (S_{\rm{px}}\ l_{\rm{depth}})^{-0.5}\sqrt{EM^{\rm{AIA}}_{ij} (\delta \langle T^{\rm{AIA}}_{ij} \rangle )^2 + \frac{(\langle T^{\rm{AIA}} _{ij}\rangle) ^2}{4EM^{\rm{AIA}}_{ij}} (\delta EM^{\rm{AIA}}_{ij})^2 }.
\end{equation}
Finally using Eq.(\ref{eqa0b}) the uncertainties on the total thermal energy (see Eq.(\ref{eq9})) of the FOV AIA data are
\begin{equation}\label{eqa4}
\delta W^{\rm{AIA}}_{\rm{therm}}=S_{\rm{px}}\ l_{\rm{depth}} \sqrt{\sum_{i=1}^{N_{\rm{px}}} \sum_{j=1}^{N_{\rm{px}}} (\delta w^{\rm{AIA}}_{ij}) ^2}.
\end{equation}

The same approach was used in the calculation of the uncertainties on the thermal energy (see Eq.(\ref{eq6})) detected by \rhessi\
\begin{equation}\label{eqa5}
\delta W^{\rm{RHESSI}}_{\rm{therm}}=3 k_B \sqrt{V} \sqrt{ EM_{\rm{RHESSI}}\delta  T_{\rm{RHESSI}} ^2 + \frac{T_{\rm{RHESSI}} ^2 }{4EM_{\rm{RHESSI}}} \delta EM_{\rm{RHESSI}}^2 },
\end{equation}
where $\delta EM_{\rm{RHESSI}}$ and $\delta  T_{\rm{RHESSI}}$  are the errors on the \rhessi\ best-fit parameters and are returned by OSPEX\footnote{\url{https://hesperia.gsfc.nasa.gov/ssw/packages/spex/doc/ospex_explanation.htm}} and based on the curvature matrix in parameter space.

\end{document}